\documentclass[letterpaper]{article}
\usepackage{aaai}
\usepackage{times}
\usepackage{helvet}
\usepackage{courier}
\usepackage{balance}  
\usepackage{paralist}      

\usepackage{graphicx}
\usepackage[usenames]{xcolor}
\usepackage{caption}
\usepackage{times}    
\usepackage{url}      
\usepackage{subfigure} 
\usepackage{longtable}
\usepackage{paralist}
\usepackage[square, comma, sort&compress, numbers]{natbib}

\makeatletter
\def\url@leostyle{%
  \@ifundefined{selectfont}{\def\UrlFont{\sf}}{\def\UrlFont{\small\bf\ttfamily}}}
\makeatother
\def\pprw{8.5in}
\def\pprh{11in}

\usepackage[pdftex,pdfpagelabels=false]{hyperref}

\makeatletter
\def\@xfootnote[#1]{%
  \protected@xdef\@thefnmark{#1}%
  \@footnotemark\@footnotetext}
\makeatother

\begin{document}
\title{Using Twitter to Understand Public Interest in Climate Change:\\ The case of Qatar}

\author{Sofiane Abbar, Tahar Zanouda, Laure Berti-Equille, Javier Borge-Holthoefer\\
Qatar Computing Research Institute\\
Hamad Bin Khalifa University\\
PO 5825, Doha, Qatar\\
\{sabbar, tzanouda, lberti, jborge\}@qf.org.qa}

\maketitle
\begin{abstract}
Climate change has received an extensive attention from public opinion in the last couple of years, after being considered for decades as an exclusive scientific debate. Governments and world-wide organizations such as the United Nations are working more than ever on raising and maintaining public awareness toward this global issue. In the present study, we examine and analyze Climate Change conversations in Qatar's Twittersphere, and sense public awareness towards this global and shared problem in general, and its various related topics in particular. Such topics include but are not limited to politics, economy, disasters, energy and sandstorms. To address this concern, we collect and analyze a large dataset of 109 million tweets posted by 98K distinct users living in Qatar -- one of the largest emitters of $co2$ worldwide. We use a taxonomy of climate change topics created as part of the United Nations Pulse project to capture the climate change discourse in more than 36K tweets. We also examine which topics people refer to when they discuss climate change, and perform different analysis to understand the temporal dynamics of public interest toward these topics.
\end{abstract}

\section{Introduction}
\label{sec:intro}

In 2011, the state of Qatar was designated by the World Bank as the highest per-capita emitters of CO2 in the world with more than 44.02 metric tons per person \cite{worldbank}. This surprisingly high rank is actually due to Qatar's massive energy sector and relatively small population. With a population of only 2.4 million in 2015 (1.8 million in 2011), Qatar was in 2013 the world's fourth-largest dry natural gas producer, and world's first liquefied natural gas (LNG) exporter in 2014 \cite{qnb}. Given the serious implications that such high emissions of greenhouse gases may have on population's health and wellbeing and on the national eco-system and biodiversity equilibrium, the government of Qatar has taken many steps toward limiting the country's ecological footprint \cite{Alhorr2014167}. For instance, Qatar hosted in 2012 the Conference of the Parties (COP18) of the United Nations Framework Convention on Climate Change where an agreement was signed to extend the life of Kyoto protocol till 2020 \cite{unfccc}. 

Qatar's mission to limit the country's carbon footprint is part of a global movement to act on the urgent warnings emanating from the scientific communities about Climate Change. Governments and world-wide organizations such as the United Nations are working more than ever on raising and maintaining public awareness toward this issue.  While sensing the engagement of governments to mitigate the causes of climate changes such as dioxide carbon emissions is relatively straightforward by looking at the international treaties and protocols ratified, it is not clear how to achieve the same objective at citizens' scale, who are part of the causes and consequences of this shared problem. Traditional surveys such as field interviews are costly and do not scale for obvious reasons. To address this concern, we analyze Twitter's content to sense the awareness of people toward climate change and investigate their interest toward its multiple topics. The recent multiplication of hazardous natural disasters around the world and the news coverage they received have helped disseminate awareness among individuals about the consequences of climate change problems, which happens to be amplified by social network platforms.

\paragraph{Main contributions.}
In this study, we are interested in sensing the public interest in climate change of people living in Qatar. This is achieved by analyzing a large collection of tweets posted between 2011 and 2016. Unlike previous studies that focused either on a limited set of climate change related keywords or on a set of geo-coded tweets (representing only 2\% of twitter traffic), our approach aims at capturing most of the discourse related to climate change under its different aspects ranging from politics and economy to energy and disasters. Our main contributions include: (i) We identified of 117K users (5\% of the entire population) claiming they live in Qatar and collected all their climate change related tweets using a rich taxonomy of topics developed by the United Nations; (ii) We propose to use network backbones analysis to better understand relationships between hashtags in co-occurrence graphs; (iii) We analyze different time series of different topics and infer the most driving topics discussed in the context of climate change.

\paragraph{Challenges.}
The main challenge we encountered in this study was identifying a relevant and accurate taxonomy of terms to use in order to retrieve climate change related posts. Many studies have focused on the usage of a single word, i.e., {\em climate} in \cite{cody2015climate}, {\em climate change} in \cite{Kirilenko2014171}, and some of them added a handful of other terms to it such as {\em global warming} in \cite{jang2015polarized,veltri2015climate}. We finally decided to use a taxonomy developed by a United Nations work group including climate change experts and social scientists \cite{ungptax}. The taxonomy is composed by a set of topics, each one is defined by a set of keywords or hashtags that should co-occur in a particular way. The experts have tried as much as possible to keep the set of terms that define topics disjoint. The other challenge was to identify correctly users living in Qatar and crawl all their tweets. We have provided a heuristic based on following relationship and a gazetteer of locations to find relevant users. Parsing a collection of 109 million tweets to extract climate change related ones, and working to keep the level of noise to its minimum possible level was a challenging task as many manual checking was required.  

\paragraph{Context of the study.}
 In this study, we focus on Twitter conversations related to environmental issues posted by people living in Qatar. Qatar is a small peninsula in the Arabian Gulf. It is the home of a native population of approximately 300,000 Qataris, who live among a total population of 2.4 million \cite{snoj2013}. Internet penetration is high in Qatar; 85\% of the population has Internet access. In addition, social media are popular and widely used among the native population  --65\% of Qataris have an Instagram account, 44\% are on Facebook, and 46\% use Twitter \cite{damian2015}.

\paragraph{Roadmap.}
The rest of the paper is organized as follows. We start by providing a literature review on related work. Then, we define the methods used for the analysis. Next, we describe the  datasets we collected. Then, follows a section that presents and discusses our findings. Finally, we conclude the paper with some insights. 

\section{Literature Review}

Climate change is a major challenge facing the world, with myriad causes and consequences. One of the leading contributors to the rising levels of CO2 emissions is the activity of humans in urban places, as a consequence, the unmanageable rising of CO2 emissions altered global climatic conditions and indirectly harms the environmental ecosystem \footnote{http://www.un.org/climatechange/}. In order to strategically raise public awareness on climate change issues, and protect the world's shared natural environment for sustainable and spiritual benefit, researchers have analyzed public opinions to identify public awareness, examine this trend on the social media sphere and investigate lasting public awareness plans. 

As most people have limited resources to get accurate data on air pollution and climate change, Wang  and Paul \cite{wang2015social} have investigated the value of social media as a sensor of air quality and public response. Authors analyzed more than 93 million messages from Sina Weibo, China\textquoteright s largest microblogging service and found a high correlation with particle pollution levels, which makes the use of social media data valuable to augment existing air pollution surveillance data, especially perception. 

Alternatively, other studies have demonstrated the utility of social media data sources to analyze public awareness about climate change, and the influence of mass media to interpret global climate problems from the different angles (causes and remedies) \cite{hart2014threat}. While Veltri and Atanasova \cite{veltri2015climate} highlighted the potential opportunity of using social media to analyze and understand public opinion dynamics. Oltreanu et al. \cite{olteanu2015comparing} compared the coverage of climate change in both, social media and mainstream news. This comparison uncovered significant differences between triggers, actions, and news values of events covered in both types of media. For instance, mainstream news sources frequently feature extreme weather events framed as being a consequence of climate change, as well as high-profile government publications and meetings. In contrast, actions by individuals, legal actions involving governments, and  investigative journalism feature frequently as viral events in social media. 

As most of people have no deep scientific understanding on climate change, their perceptions are most likely influenced by framing strategies \cite{weber2011public} , which leads the research community to question public opinions over time and location, along with the role  that mass media played. Cody et al. \cite{cody2015climate}  studied user sentiment toward climate change by analyzing environment-related tweets posted between 2008 and 2014. The study shed a light on how climate news, events, and disasters affects people's sentiment. An et al. \cite{an2014tracking}  also conduct an opinion analysis of climate change related tweets. Authors aimed at understanding whether Twitter data mining can complement and supplement insights about climate change perceptions, especially how such perceptions may change over time upon exposure to climate related hazards. A combination of techniques drawn from text mining, hierarchical sentiment analysis and time series methods is employed for this purpose. At a large scale, Jang and Hart \cite{jang2015polarized} investigated how people incorporate climate change into everyday conversation by the stream of Twitter over two years. Their study shows geographical differences in the use of frames and terms concerning climate change, and how certain frames that promote skepticism about climate change are widely circulated by users within specific regional and political contexts. 

In this paper, we analyze climate change conversations and air quality trends in Qatar through the lens of social media data, and investigate the ability of measuring public awareness, along with the impact of environmental factors on public conversations.


\section{Methods}
\label{sec:methods}

We describe in this section the methods we used in this study.

\subsection{Building the Network of Co-occurrences}
\label{subsec:co-occurrence}
The relationship between different concepts in the dataset can be suitably represented as a network.  Nodes in the network represent concepts, and a link from node $i$ to node $j$ implies that the corresponding concepts of the two nodes co-occur in the same tweet.  More importantly, each node bears an associated quantity $a$ that stands for the proportion of tweets in which it appears. On the other hand, each link from $i$ to $j$ has an associated weight $(i,j,w)$, which states the fraction of tweets in which these two nodes co-occur.

\subsection{Network backbone extraction}
\label{subsec:backbone}
Networks can be described from different levels of analysis. At the {\em micro} level, the focus lies on single nodes and their specific positions within the overall structure; this level can be described in terms of node degree, strength or clustering coefficient, among other metrics. At the {\em macro} level, the focus shifts to the aggregation of those metrics and the properties of their distribution. Between these two extremes, we have a third level of analysis, the {\em mesoscale}, which aims at accounting for the complexity of networks between the position of individual nodes and the relational properties of the collectives they form. It is at this level where reduction techniques like backbone extraction operate.

Network backbone extraction refers to the filtering techniques aimed at uncovering  relevant information; in general, such techniques aim at pruning the links of a network, keeping only those which are statistically relevant. Ideally, the reduced structure is computationally more tractable while it retains most of the interesting features of the original one. We apply the state of the art backbone extraction algorithm proposed in \cite{serrano2009extracting} to the network of hashtag co-occurrences. In their work, Serrano et al. exploit the empirical trend by which link weights are heavily fluctuating, i.e., only few links carry the largest proportion of the node's strength.

\subsection{Linear Correlation}

We use Pearson product-moment coefficient to measure the linear correlation between different pairs of variables (environment topics and weather conditions) of interest in this study. Pearson correlation score ranges in $[-1,1]$ where 1 (resp. -1) means the existence of total positive (resp. negative) correlation and 0 means no correlation at all. Pearson formula between two time series X and Y is given below:

\begin{equation}
 \rho_{X,Y} = \frac{E[(X-\mu_{X})(Y-\mu_{Y})]}{\sigma_{X} \sigma_{Y}}
\end{equation}

where: $\mu_{X}, \mu_{Y}$ are the means of $X,Y$ respectively, and $\sigma_{X}, \sigma_{Y}$ are the standard deviations of the two corresponding time series.

\section{Datasets}
\label{sec:data}

To conduct this study, we need to combine data from two different sources: Twitter and Weather Underground.

We use Twitter to collect user generated content about environment related topics. As there is not solicitation for users to post on Twitter, we claim that this data reflects the actual and natural responses to users toward different environmental topics as they get exposed to them through news media or witness them happening during natural disasters, heavy rains, drought, etc.  Underground Weather is used to collect daily weather conditions in Doha, the capital of Qatar. It is worth  noticing that Qatar is a relatively small country (area: 11,586 $km^{2}$) with more than 80\% of its population based in the city of Doha. Thus, the weather does not vary that much from a location to another in the country.
We explain in the following the processes by which we collected and curated the different datasets.

\subsection{Identifying users of Qatar}
First, we created a gazetteer of locations in Qatar including the names of main cities and districts in both English and Arabic, and under different spellings. We matched this list to a 45 day sample of the Twitter Decahose (a sample of 10\% of all Twitter traffic.) For each user whose tweet was captured, we requested the list of their followers based on the intuition that for an average user, followers are more likely to be from the same country. 

 Then, we requested the profiles of the followers and filter  out those users who did not mention a location corresponding to Qatar's gazetteer. The same process was iteratively executed on the list of newly added users; i.e., find their follower IDs, then follower profiles, then filter out followers with irrelevant locations, until no new users are identified. To expand the list of users even further, a social analytics tool (FollowerWonk.com) was used to obtain lists of users who mention a location associated with our gazetteer. This list was then merged with the list obtained earlier from Decahose and Twitter API. The resulting list consists of over 117K users who claim to be living in Qatar in their Twitter profiles. 

\begin{figure}
\centering
\subfigure[Tweets\label{fig:geo_coded_env_tweets}]
{\includegraphics[width=.40\columnwidth]{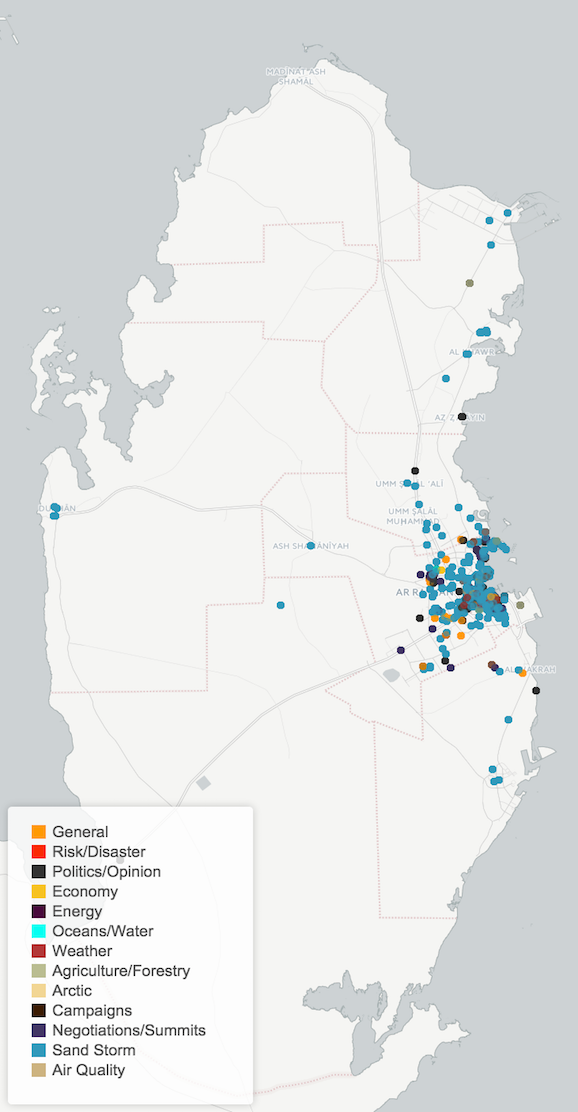}}
\subfigure[Census 2010\label{fig:map_census}]
{\includegraphics[width=.405\columnwidth]{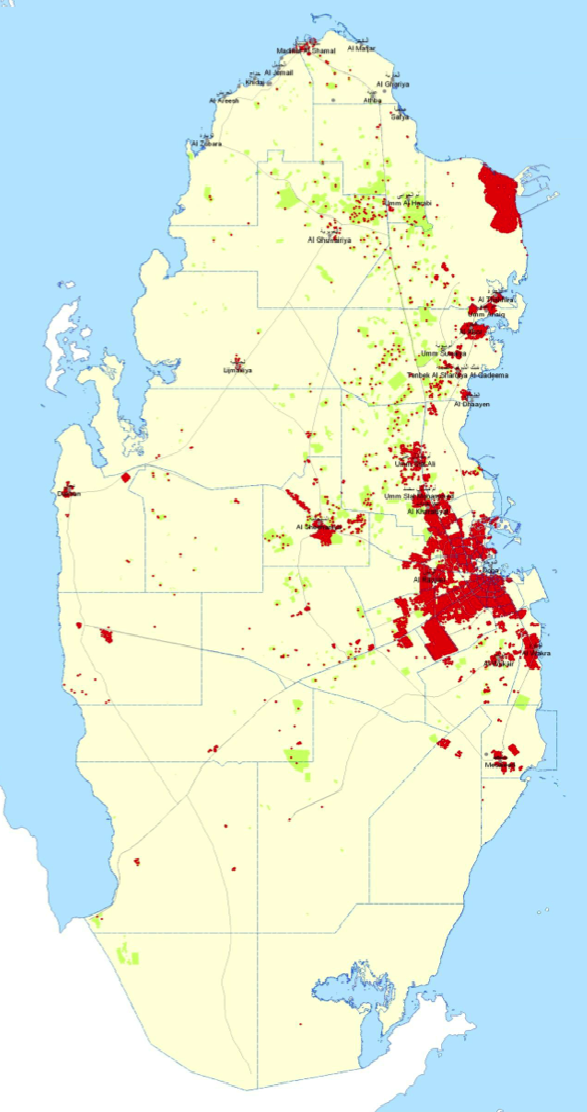}}
\caption{Qatar Maps}
\label{fig:qatar_maps}
\end{figure}

\subsection{Collecting user tweets}
We use Twitter Historical API\footnote{https://dev.twitter.com/, last accessed on Feb,2016} to collect up to 3,200 tweets posted (the maximum allowed by Twitter API) for each of the users of Qatar identified in the previous step. This collection has been conducted twice. The first time was in July 2014 and the second time was in February 2016. Hence, we could obtain for each user up to 6,400 tweets he posted over a period of nine years (from 2007 to 2016). This collection generated a total of 109.6 million tweets posted by more than 98,066 different users. The remaining missing users have either a disabled or protected account. As Twitter was launched in 2006 only, the activity of the users of Qatar was very marginal before 2011. Thus, we focus in this study on the time period spanning from January 1rst, 2011 to January 1rst, 2016. 

\subsection{Weather data}
We use Weather Underground (WU) API \cite{wu} to collect data about daily weather conditions in Qatar for the same time period. Out of the sixty different measurements WU API returns, we were particularly interested in the ones that are more likely to sprinkle climate change related tweets, these measurements include: temperature ($^{\circ}\mathrm{C}$), humidity (\%), wind speed (km/h), visibility (km), and precipitation (mm) (including: rain, snow, etc.)

\subsection{Extracting environment and climate change tweets}

In order to identify the climate change related content, we use a taxonomy of keywords grouped into topics on interest.  Previous studies have used  simple keywords to filter climate change tweets.  Cody and Reagan for instance used only one word {\em climate} \cite{cody2015climate}. Veltri and Atanasova used two keywords {\em climate change} and {\em global warming} \cite{veltri2015climate} while  Andrei and Svetlana added ​translations of the aforementioned words in German, Portuguese, Russian and Spanish \cite{Kirilenko2014171}. Jang and Har also used the keywords {\em climate change} and  {\em  global warming} paired with some specific terms related to different frames~\cite{jang2015polarized}. 

We use in this paper a taxonomy of climate change topics\cite{ungptax} developed by the UN Global Pulse team. The taxonomy consists of set of topics, each of which is defined with a list of terms and hashtags that should co-occur in a particular way. According to Ren\'e Clausen Nielsen -- data innovation specialist in UN Global Pulse -- the taxonomy is created by climate change experts as follows: (1) find a list of relevant words. (2) find synonyms, plurals, and abbreviations by searching Twitter and dictionaries. (3) Group words into distinct topics. (4) Create negative filters to get rid of occurring noise~\cite{reneclausen2016}. 

An important feature of the used taxonomy is the fact that topics are carefully defined not to allow overlapping.  These topics are: {\em General}, {\em Risk\&Disaster} ,{\em Politics\&Opinion}, {\em Economy}, {\em Energy}, {\em Oceans\&Water} ,{\em Weather} ,{\em Agriculture\&Forestry},  {\em Arctic} , {\em Campaigns} , {\em  Negotiations\&Summits}. For a tweet to be labeled as belonging to a particular topic, it should satisfy a list of conditions. For instance, a tweet is about {\em Politics\&Opinions} if and only if it {\em CONTAINS ANY} \{climate, enviro, environment, carbon\} {\em AND NOT CONTAINS ANY} \{monoxide, fork, copy, fiber\} {\em AND CONTAINS ANY} \{politics, group of eight, \ldots \}.

The dusty construction sites in the middle of the city, and the flat dusty landscape stretching out of the capital's borders, both contribute to complicated air quality conditions. We added  {\em Air Quality} topic separately to captures conversations around air quality and sense how people talk about the air they breath by tracking words that refer to air and air pollution (e.g.,  air quality, air pollution ). 
Like Countries in the region, Qatar has built cities in the middle of the desert. Heavy sandstorms hit the country regularly, and keep fine dust hanging in the air for days. Thus we added  {\em Sandstorm} topic composed of terms such as duststorm, dusty, and sand storm. A full list of topics with their respective definitions, terms, and matching conditions can be downloaded from~\cite{ourtaxonomy}.

\subsection{Data Curation}
Next, we started doing a qualitative analysis on our dataset to determine content that might not be relevant in our analysis. We have noticed that some irrelevant keywords such as {\em Syria, California, Colorado, etc.} were appearing to the top of co-occurrence lists. The manual inspection allowed us to identify some bots that were re-tweeting all tweets about Syria with by appending US state names as hashtags.  

At the end of this process, we obtained a clean set of 36,612 environment relevant tweets posted by 8,470 different users (also known as tweeps).

Figure~\ref{fig:qatar_maps} shows the geo-coded tweets related to environment that we have gathered, and a population census from 2010 for comparison. The most populated area is definitely around the capital city of Doha, with beaches and main roads being other popular posting sites. A shallow visual inspection reveals that posting proficiency is correlated with population density in general. 

\begin{table}[ht]
\caption{Description of the final set of topics related to environment and climate changes with some basic statistics.}
\label{tbl:terms}
\small
\begin{tabular}{c||c|c|c|c}
Category  & \#Tweets & \#Users & T/U & \%T \tabularnewline
\hline 
\hline 
General  &  3828   & 1293 & 2.96 & 10\% \tabularnewline
\hline 
Risk/Disaster  & 320 & 129 & 2.48 & 8\% \tabularnewline
\hline 
Politics/Opinion  & 11124  & 3192 & 3.48 & 30\%\tabularnewline
\hline 
Economy  & 2615 & 1038 & 2.51 & 07\% \tabularnewline
\hline 
Negotiations/Summits  & 7576  & 1080 & \textbf{7.01} & 20\% \tabularnewline
\hline 
Energy  & 1716 & 722 & 2.37 & 4\% \tabularnewline
\hline
Agriculture/Forestry & 1119 & 659 & 1.69 & 3\% \tabularnewline
\hline
Ocean/Water & 458 & 289 & 1.58 & 1\% \tabularnewline
\hline
Campaigns & 470 & 159 & 2.95 & 1\% \tabularnewline
\hline
Arctic & 132 & 112 & 1.17 & 0.3\% \tabularnewline
\hline
Air quality & 69 & 41 & 1.68 & 0.1\% \tabularnewline
\hline
Sandstorm & \textbf{13124} & \textbf{5605} & 2.34 & 30\% \tabularnewline
\hline
Weather  & 1139 & 793 & 1.43 & 0.03\% \tabularnewline
\hline 
\end{tabular}
\end{table}

Table \ref{tbl:terms} illustrates the final set of topics related to environment and climate changes with some basic statistics in terms of number of tweets, number of users, ratio of tweets per user and percentage of tweets related to each category.

\section{Results}
\label{sec:content}

In this section, we present the results of different analysis performed on the set of environment-related tweets described in the previous section. First, we show the temporal trends of people reactions to different environmental topics. We show also the most prominent hashtags associated with different topics. Then, we look at the co-occurrence graph of these hashtags to unveil different hashtag relationships and graph topologies. Finally, we analyze user profiles to check whether users posting about environment demonstrate any specific characteristics. 

\subsection{Temporal dynamics}
\begin{figure}[th]
\centering
  \subfigure[Tweets\label{fig:tweets_all_env}]{\includegraphics[width=.85\columnwidth]{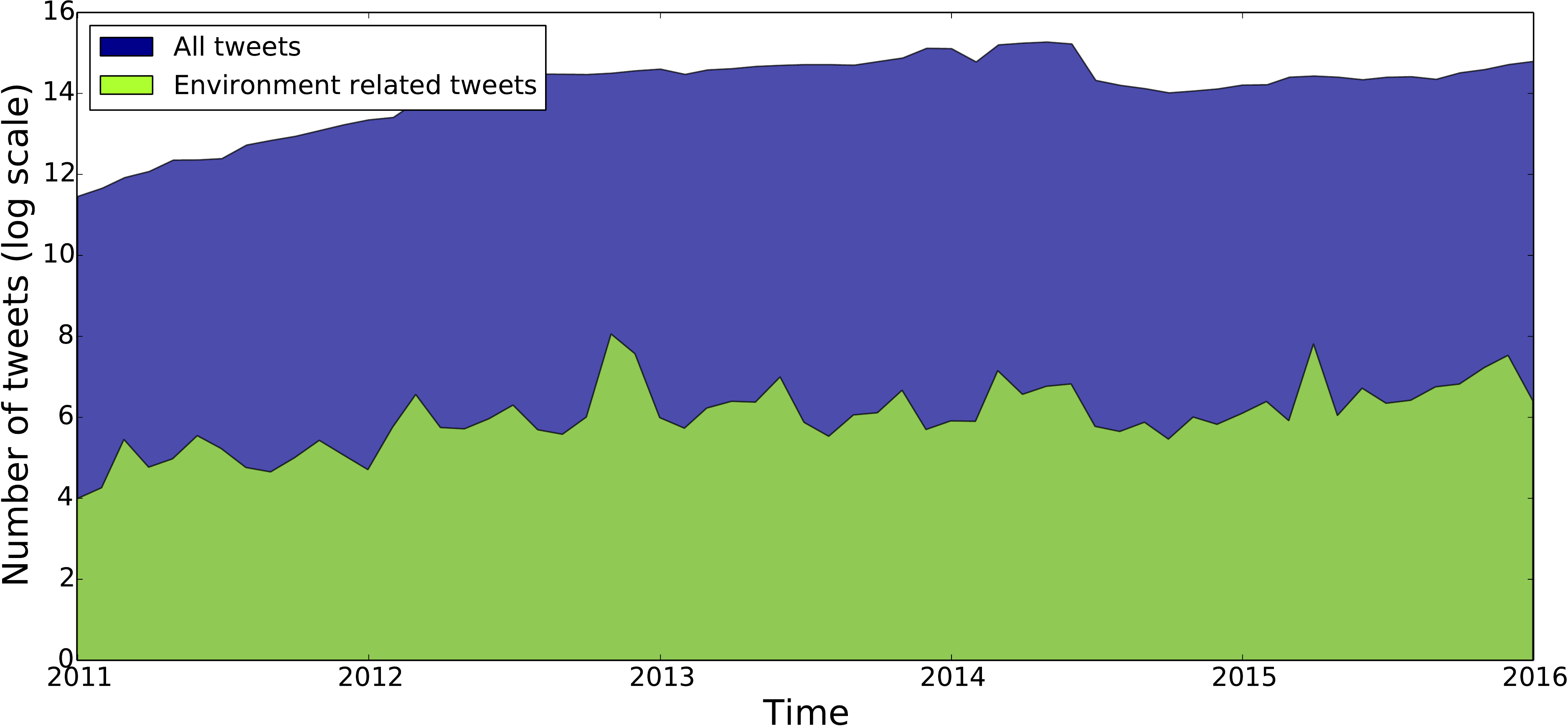}}
  \subfigure[Users\label{fig:tweeps_all_env}]{\includegraphics[width=.85\columnwidth]{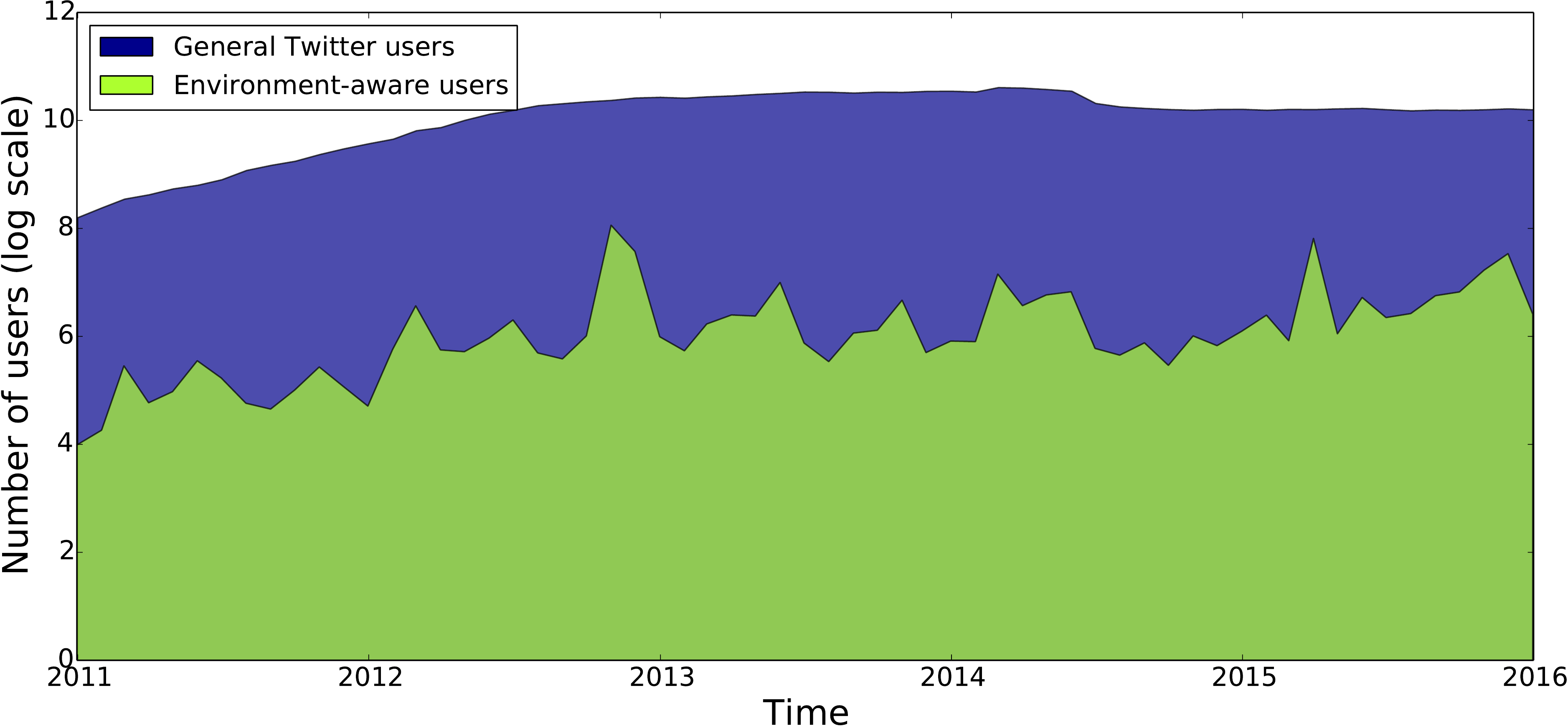}}
  \caption{Monthly volume of the total number of tweets and unique users in log scale in Qatar from January 2011 to January 2016. We see that climate change curves present a noticeable variance over time unlike the general smoother Twitter pattern.}
  \label{fig:global_trends}
\end{figure}

Figure~\ref{fig:global_trends} shows the trends of the monthly total number of tweets (a) and unique users (b) observed in the Qatar ``tweetosphere''. The blue curves represent the trends of all the tweets posted while the green curves represent environment related tweets only. To ease the reading, the y-axis is plotted in log scale. The first striking observation is that even in log scale, users' activity around environmental topics fluctuates a lot compared to the smooth general discussions on Twitter. Most of the spikes observed in Fig~\ref{fig:tweets_all_env} can be explained by looking at the tweets posted around the spike period. For instance, one of the most noticeable burst, the one happening in November-December 2012 is a response to the Conference of the Parties (COP18) that took place at Qatar late in 2012. That conference to which more than 50 presidents from around the world took part received a large news coverage locally and internationally. Yet, people living in the small peninsula of Qatar have extensively experienced the conference as it has directly impacted their everyday life in terms of commute deviations, security enforcement, and access limitations to some hotels. All this has generated an unprecedented excitement on Twitter for the event. This observation is corroborated by Fig~\ref{fig:tweeps_all_env} in which it is easy to see the important burst of the number of unique users who took part of discussions and debates around COP18 conference. The second sharper spike that takes place around March 2015 is due to a sudden strong sandstorm that happened in Qatar in that same period, causing schools and institutions to close doors on that day as it was almost impossible to drive. 
As an example, the below-mentioned tweet is posted by Qatar's Ministry of Interior on 1st April 2015 \footnote{\url{https://twitter.com/MOI_QatarEn/status/583359974278967296}}: ``{\em Visibility is almost zero in all roads inside and on highways. Motorists are requested to follow safety instructions. \#MoI\_Qatar.}''

\begin{figure*}[th]
  \centering
  \includegraphics[width=0.91\linewidth]{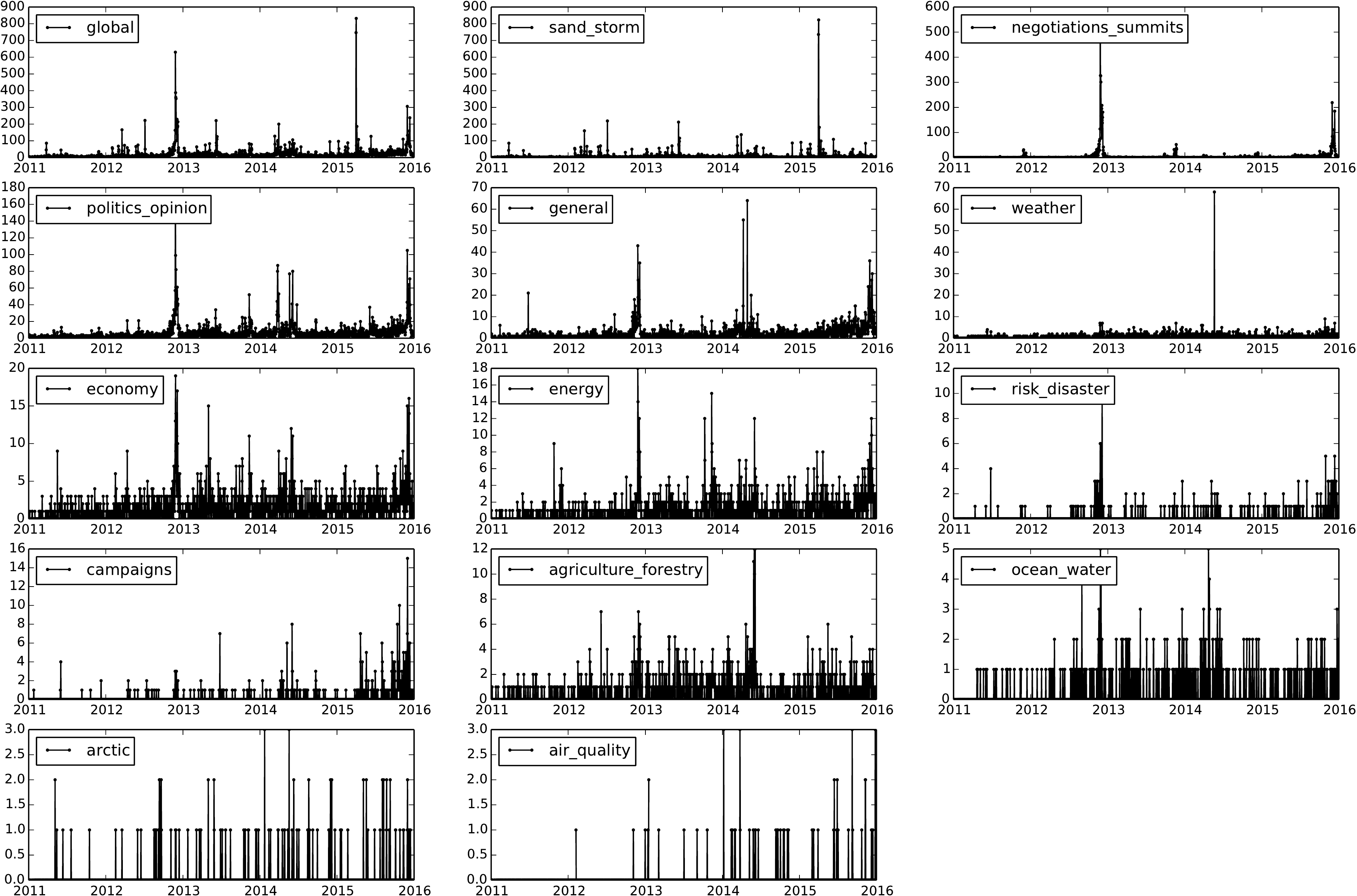}
  \caption{Daily distributions of the total number of tweets for different climate change related topics ordered by volume from top left to bottom right. Notice that y-axises are not normalized across the figures.}
  \label{fig:env_ts_topics}
\end{figure*}

Continuing further with the temporal analysis of our set of tweets, we plot in Figure~\ref{fig:env_ts_topics} the breakdown of this trend into different topics by providing the daily time series of the number of tweets  associated with each topic. The first figure to the left (Global) plots the daily time series of all environment-related tweets. Some observations can be done here. First, the different topics vary a lot in terms of volume and trends. As highlighted in Table~\ref{tbl:terms}, the {\em sandstorm} topic seems to overtake all other topics. It is clear that this topic constitute the most discussed environment related topic on Twitter by the people of Qatar as this has a direct and immediate impact on their lives. The figure shows also that when the people of Qatar discuss about environment on Twitter, it is mostly about {Politics\&Opiniom, Negotiations\&Summits}, and climate change in general. The less covered topics include {\em Oceans\&Water, Arctic} and surprisingly {\em Air Quality}. Second, one can easily see that topic time series follow different trends. Indeed, while some topics such as {\em Economic} and {\em Energy} maintain a decent level of activity throughout the years, other topics such as {\em Negotiations\&Summits} and {\em Arctic} are sparse and only acquire users' attention when a related event happens.

\subsection{Content Analysis}
\begin{figure}[h]
  \centering
  \includegraphics[width=0.99\linewidth]{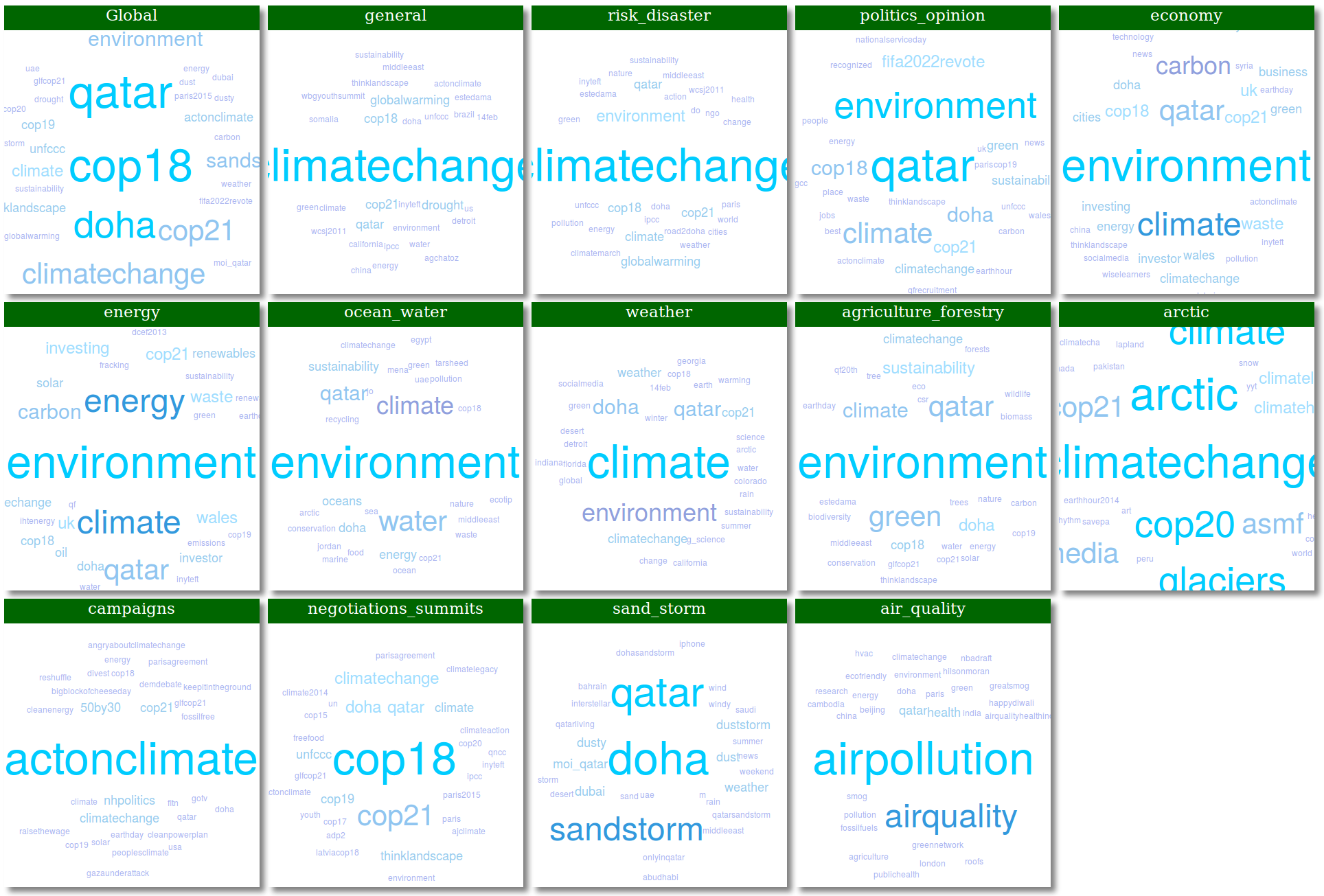}
  \caption{Word clouds of hashtags corresponding to different environment-related topics. The first word cloud to the left is generated from all environment-related tweets.}
  \label{fig:env_tagclouds}
\end{figure}

In order to visually inspect the content of tweets in different topics, we plot Figure~\ref{fig:env_tagclouds} in  for each topic, we generate its word cloud of the most frequent hashtags mentioned in the corresponding tweets. The font size of hashtags is proportional to their frequencies in different topics. 
These word clouds reveal for instance that the hashtag {\em \#COP18} is dominating in our dataset of tweets (global word cloud) and is obviously dominating in tweets belonging to {\em Negotiations\&Summits}. 
We also found the hashtags {\em \#environment, \#climate}, and {\em \#carbon} to be dominating in {\em Economy} and {\em Energy} as {\em \#carbon} is relevant to both topics (the carbon economy). {\em \#environment} dominates also {\em Ocean\&Water} and {\em Agriculture\&Forestry}, {\em \#climate} dominates in {\em Weather}, {\em \#actonclimate} dominates in {\em Campaigns}, {\em \#climatechange} dominates in {\em Arctic}, and {\em \#airpollution} dominates in {\em Air Quality}. 

\begin{figure*}
\centering
\subfigure[Global\label{fig:global}]{\includegraphics[width=.39\linewidth]{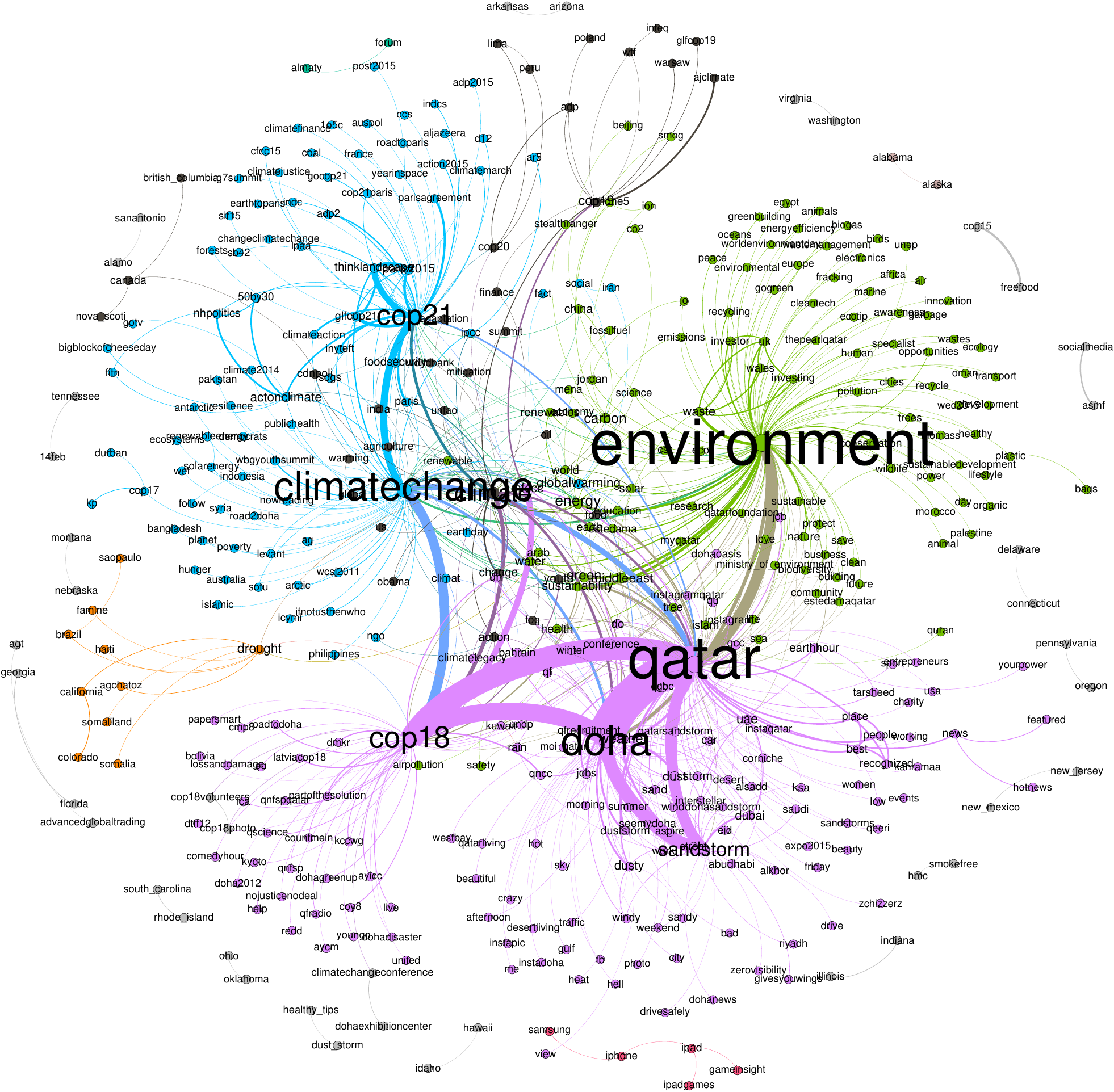}}
\hfill
\subfigure[General\label{fig:general}]{\includegraphics[width=.39\linewidth]{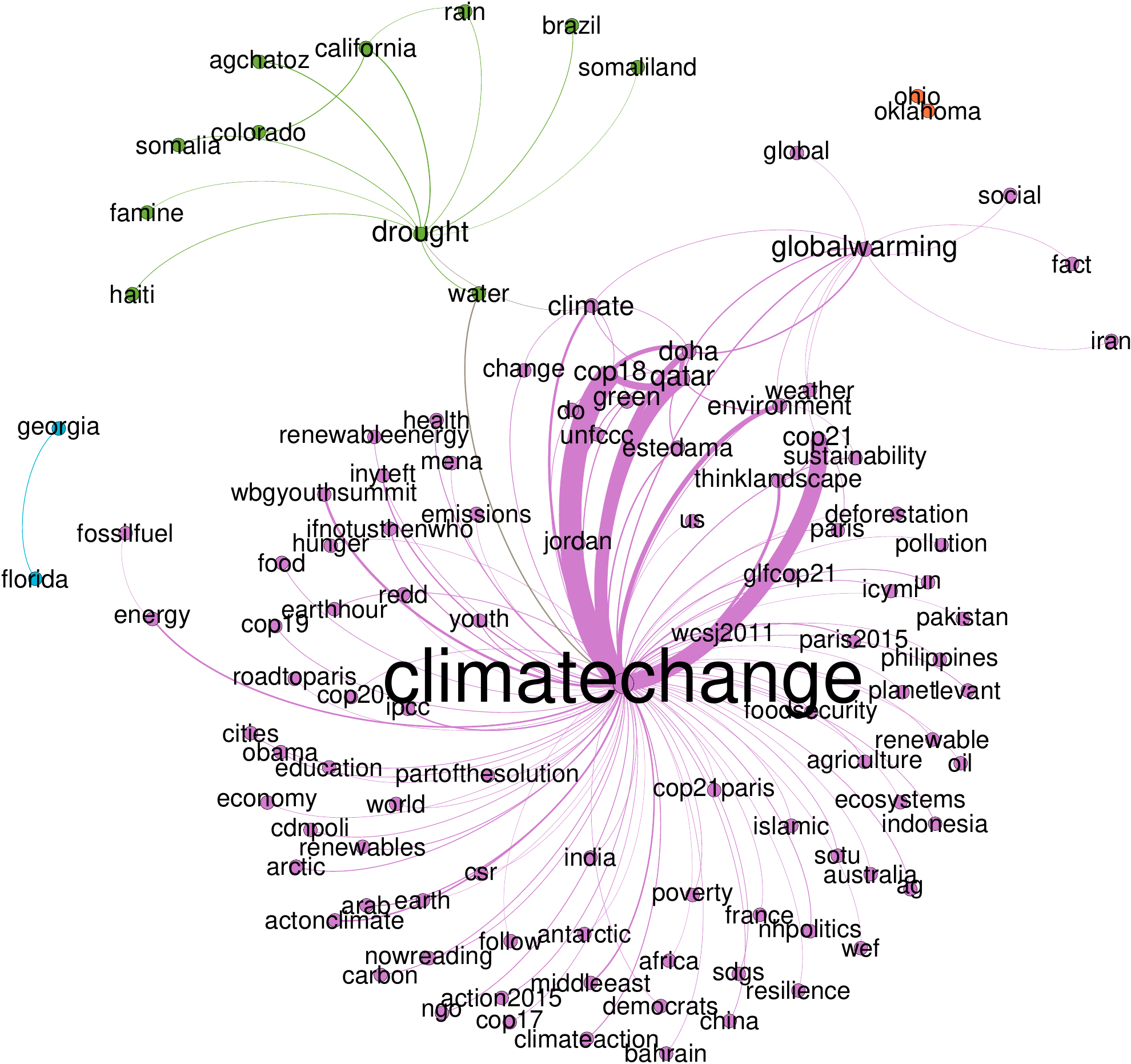}}
\hfill
\subfigure[Risk\&Disaster\label{fig:disaster}]{\includegraphics[width=.2\linewidth]{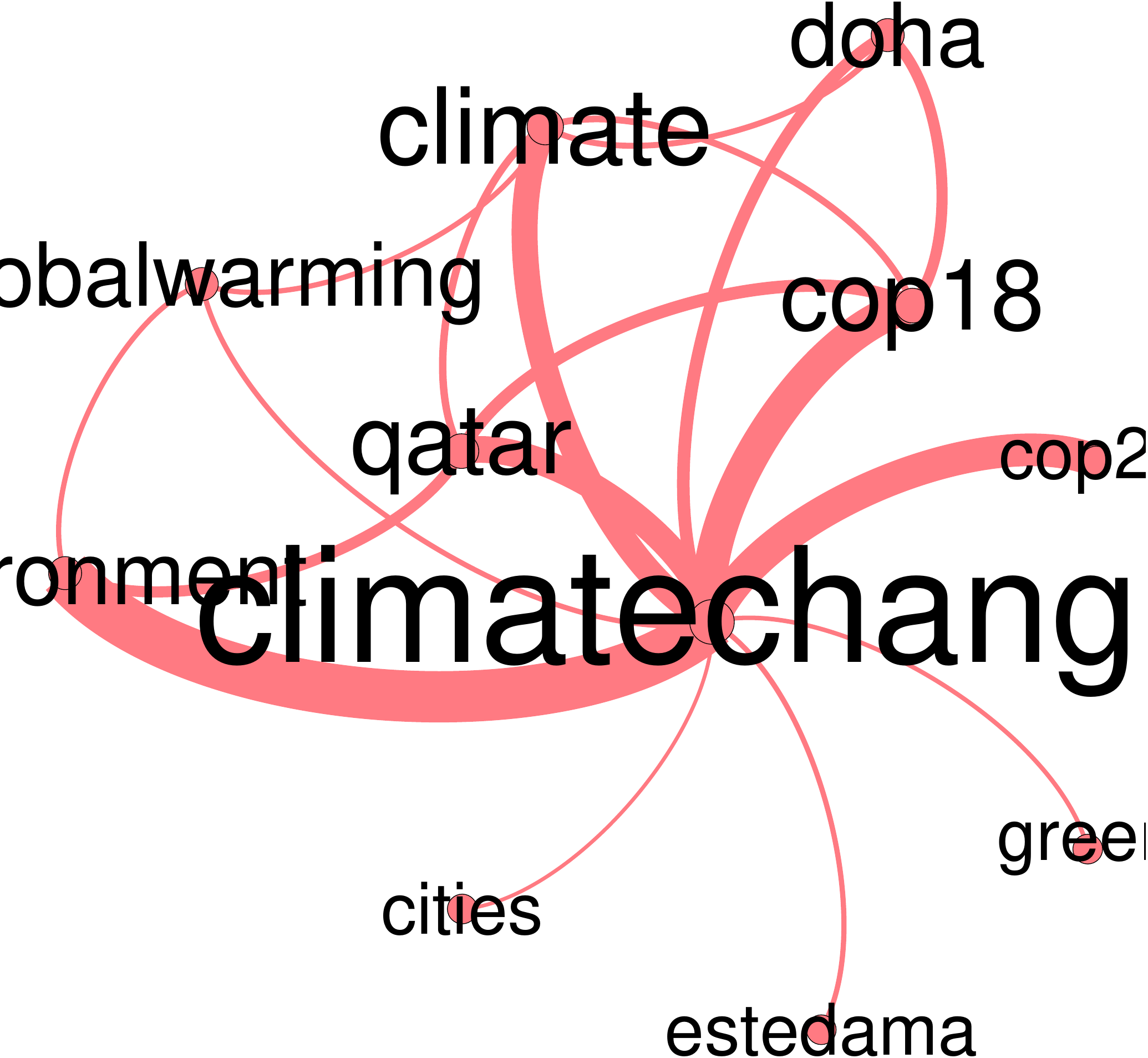}}
\subfigure[Summits\&Negotiations\label{fig:summit}]{\includegraphics[width=.33\linewidth]{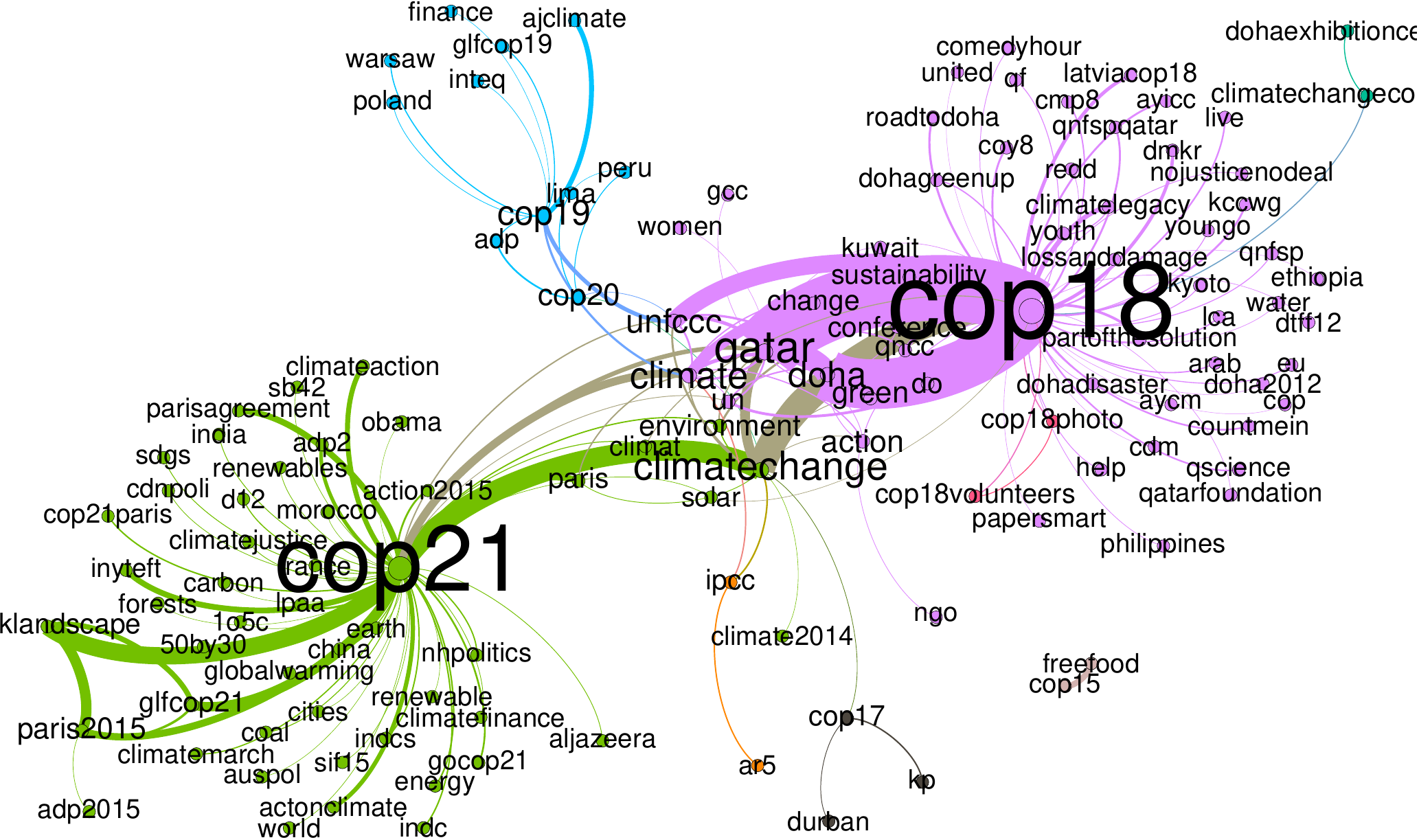}}
\hfill
\subfigure[campaigns\label{fig:campaigns}]{\includegraphics[width=.2\linewidth]{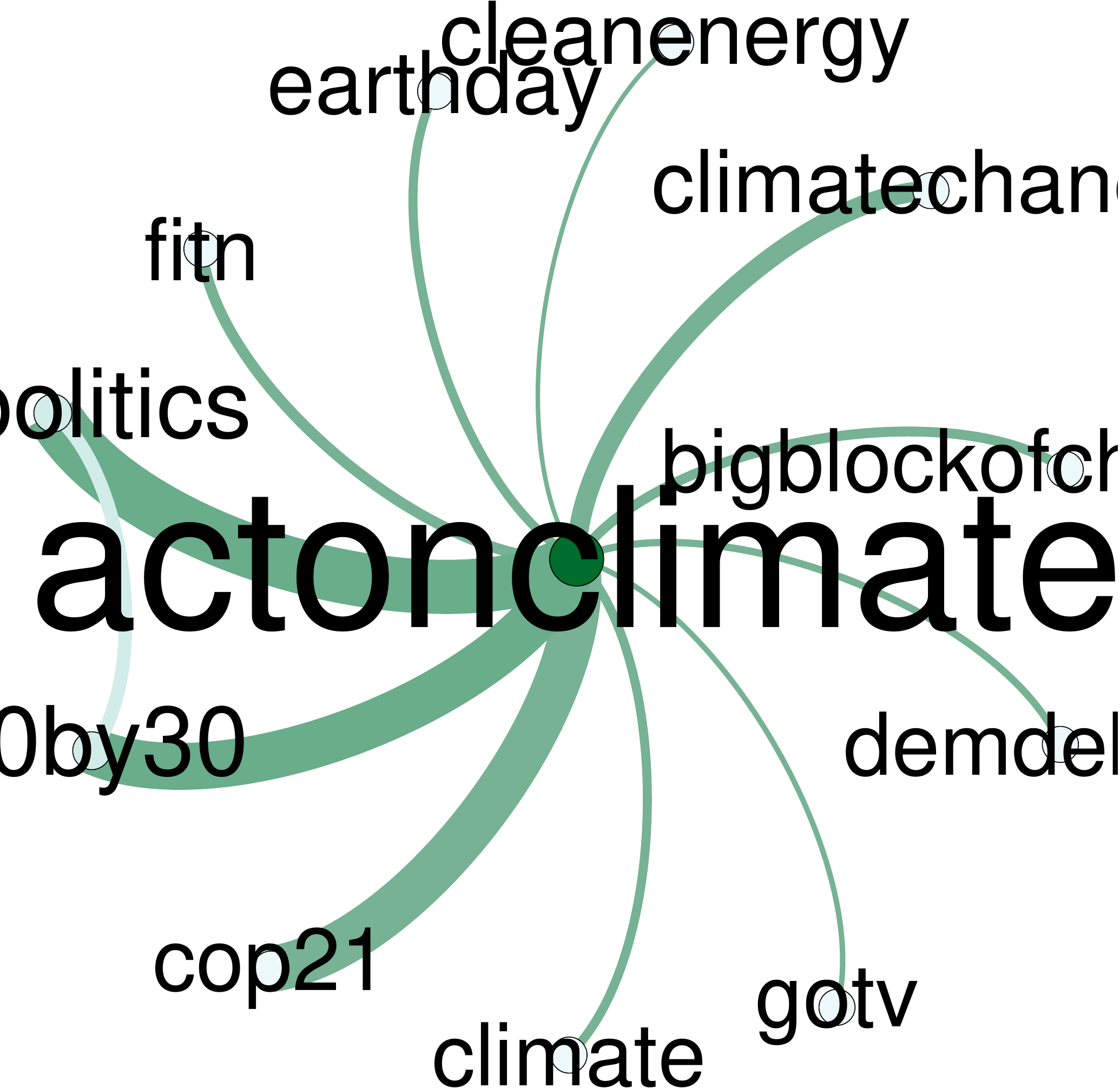}}
\hfill
\subfigure[Politics\&Opinion\label{fig:politics}]{\includegraphics[width=.33\linewidth]{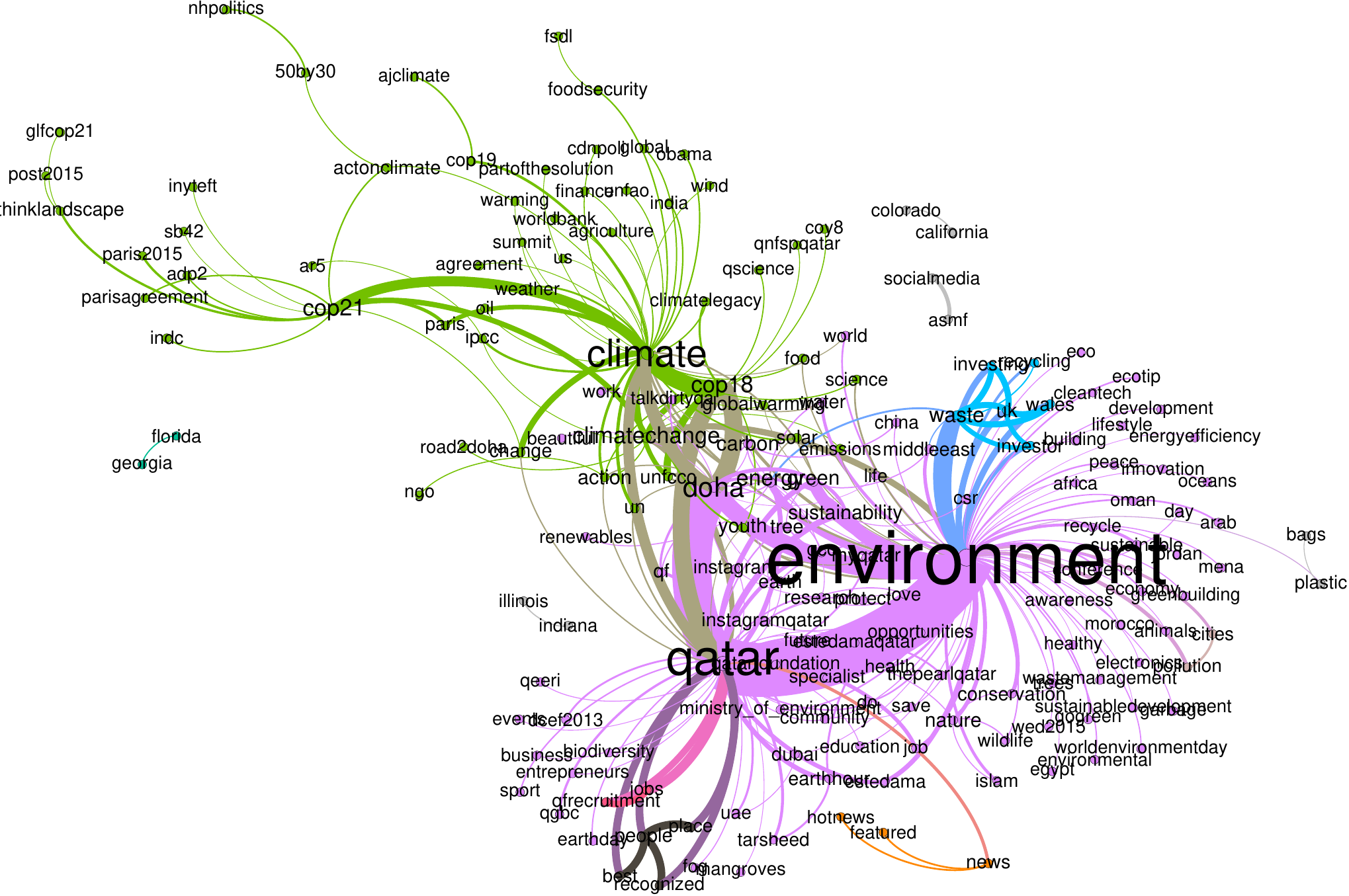}}
\subfigure[Sandstorm\label{fig:sandstorm}]{\includegraphics[width=.3\linewidth]{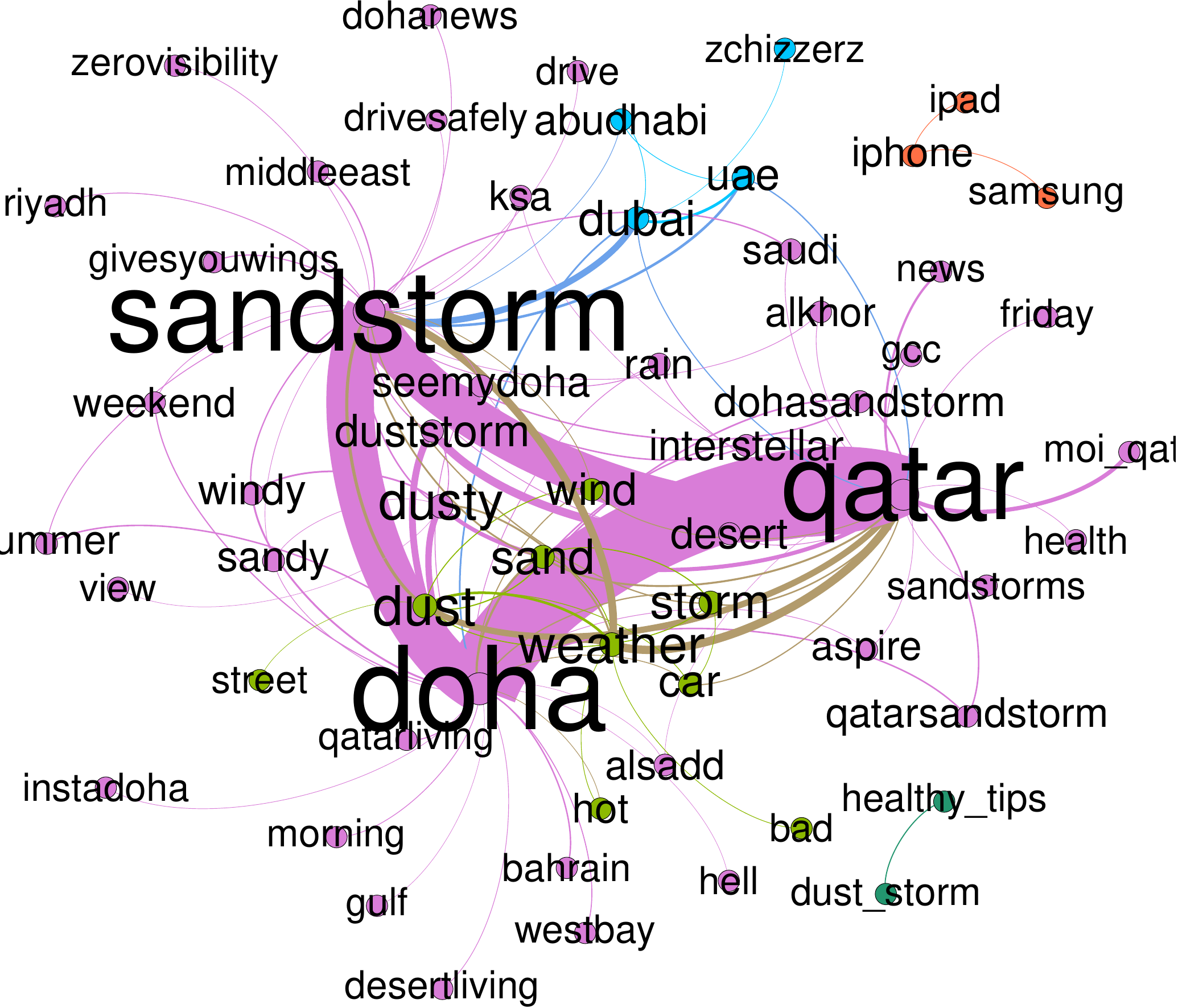}}
\hfill
\subfigure[Energy\label{fig:energy}]{\includegraphics[width=.3\linewidth]{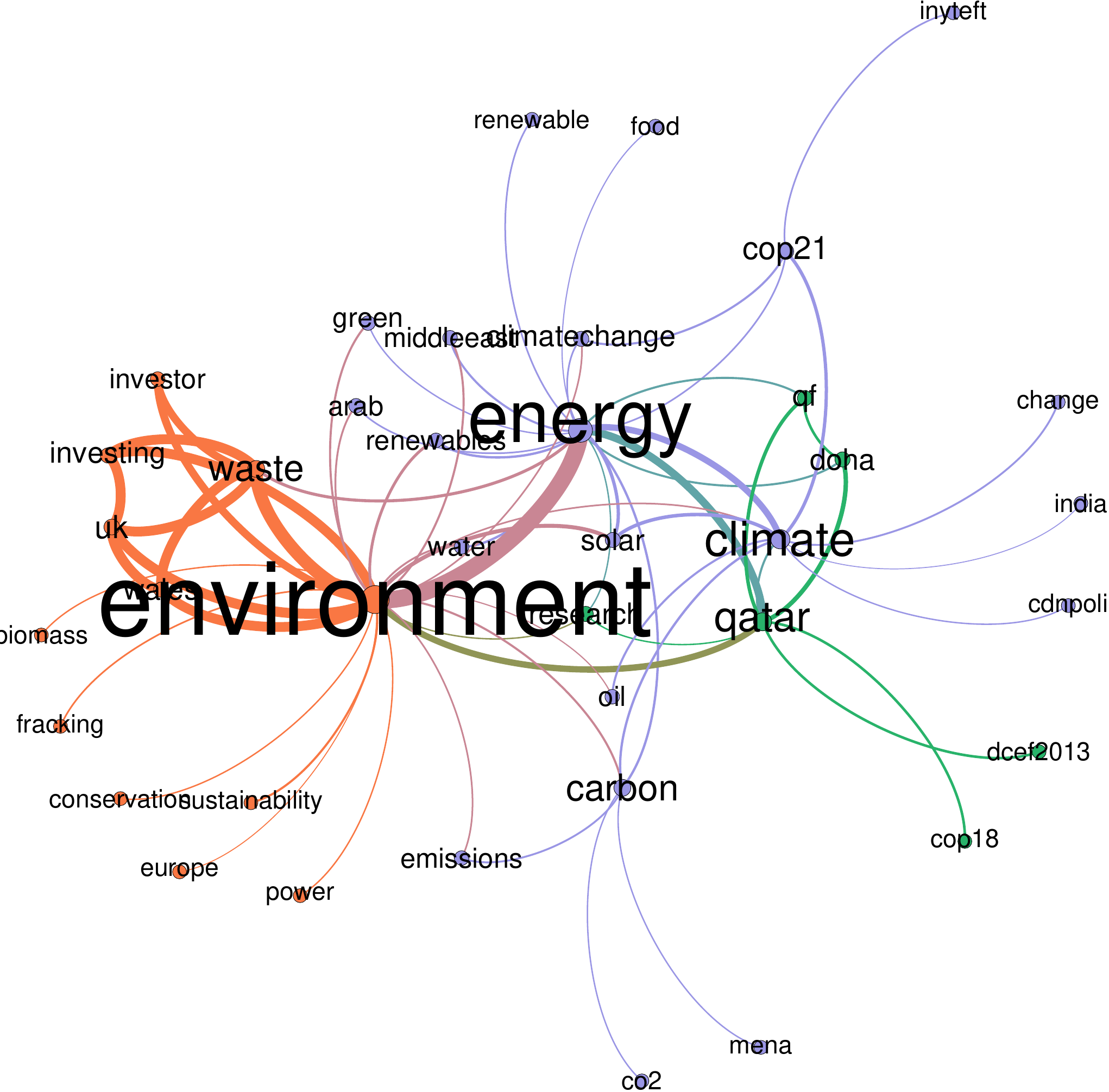}}
\hfill
\subfigure[Economy\label{fig:economy}]{\includegraphics[width=.3\linewidth]{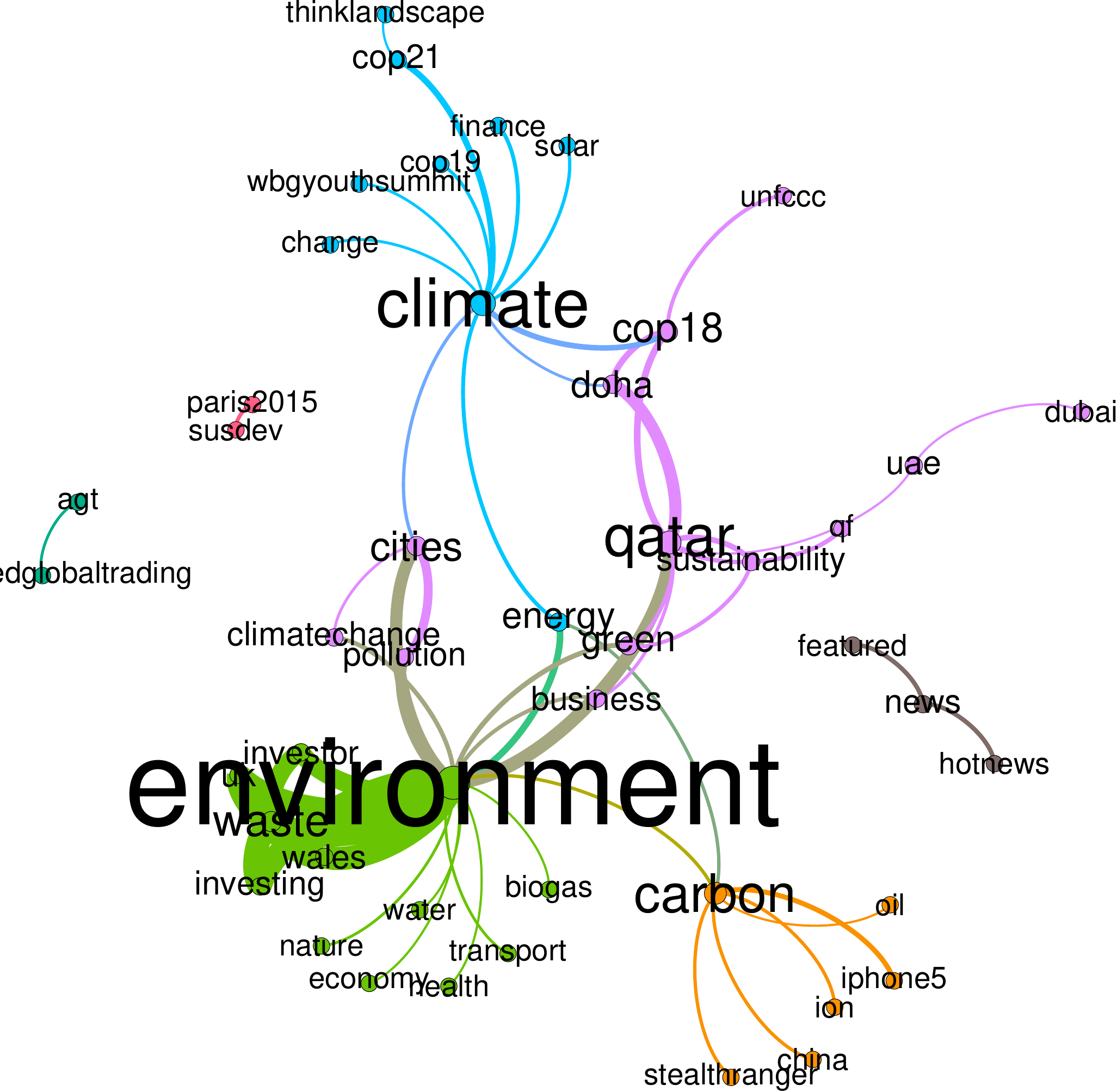}}
\subfigure[Agriculture\&Forestry\label{fig:agriculture}]{\includegraphics[width=.25\linewidth]{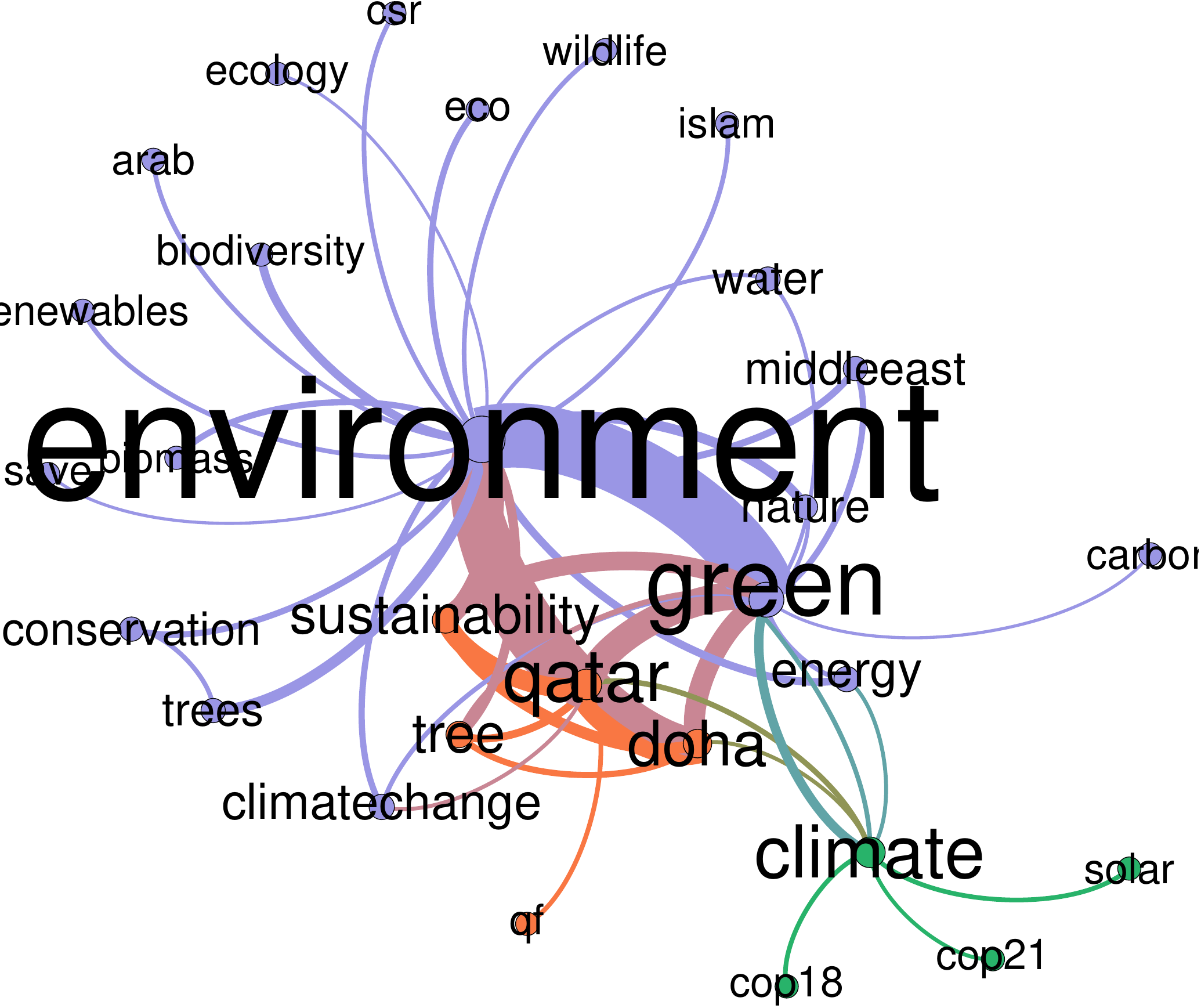}}
\hfill
\subfigure[Oceans\&Water\label{fig:oceans}]{\includegraphics[width=.22\linewidth]{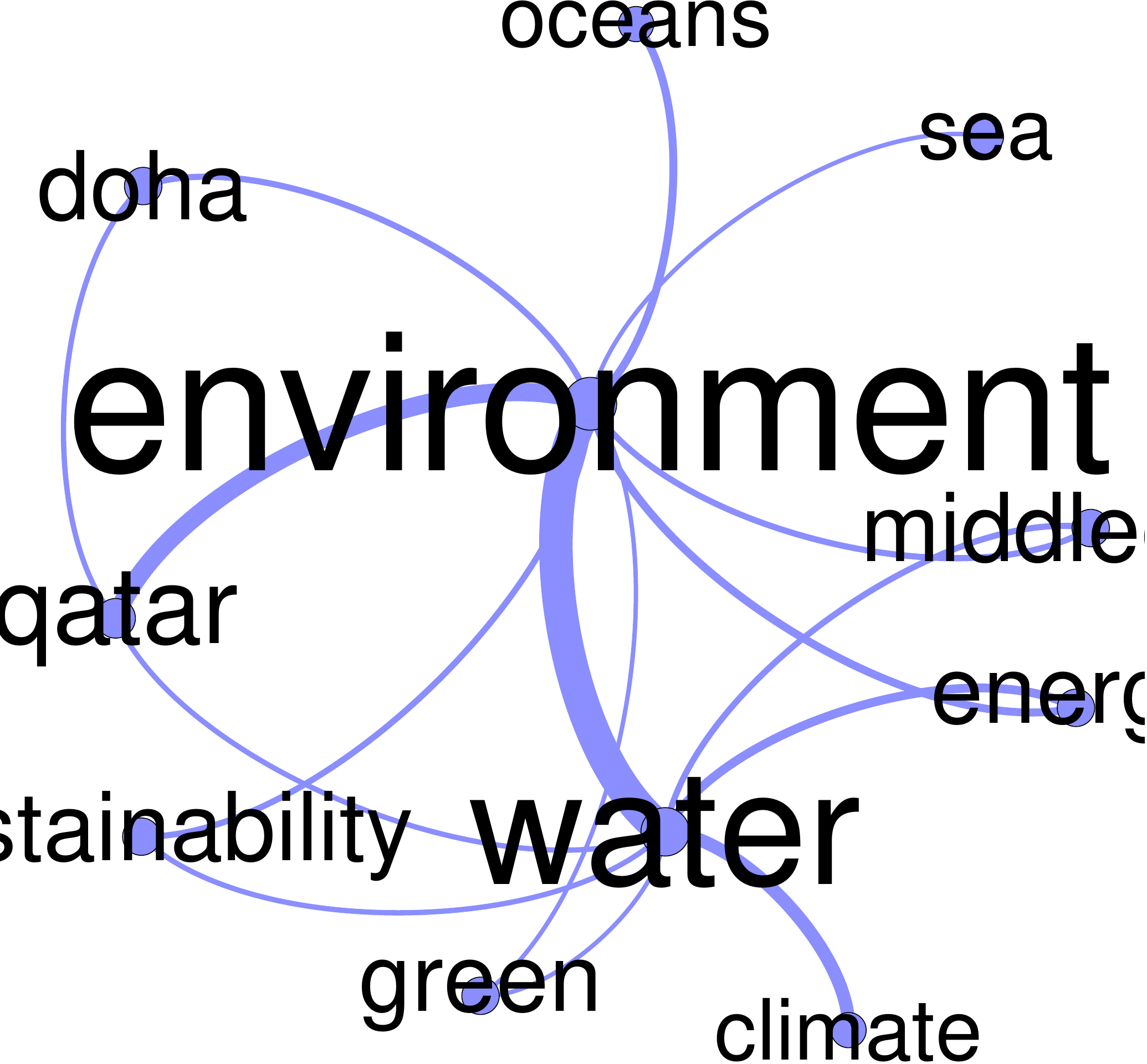}}
\hfill
\subfigure[Weather\label{fig:weather}]{\includegraphics[width=.25\linewidth]{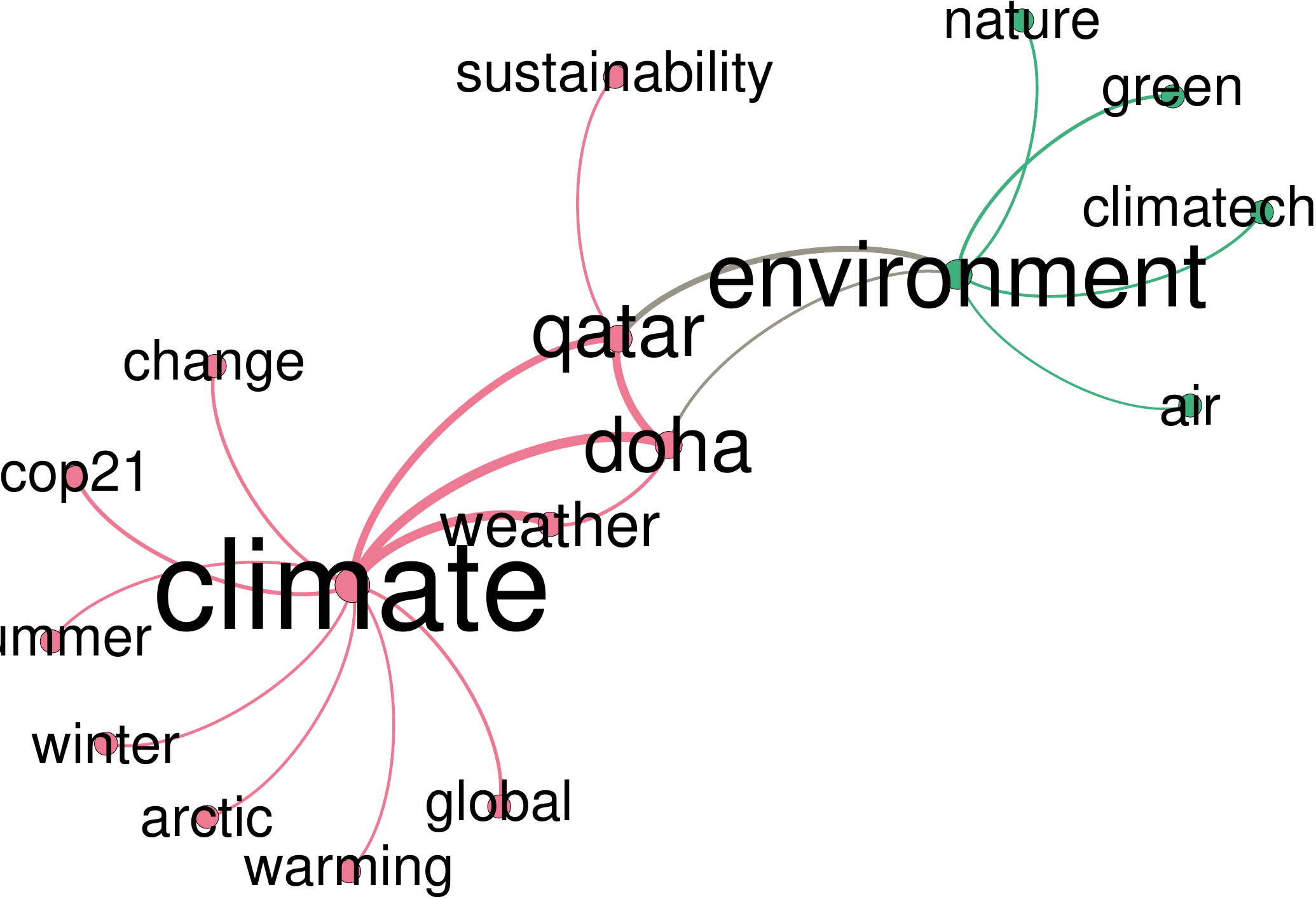}}
\caption{Graphs representing hashtag co-occurrence backbones of different climate change topics (b) through (l). (a) is the global backbone graph computed from the whole set of climate change related tweets.}
\label{fig:coocc}
\end{figure*}

To further analyze the content, we focused on the relationships between  hashtags within  topics and we built the co-occurrence networks of hashtags for the tweets of each topic. Because the generated networks were dense, it was not possible to perform any kind of visual observations on them. Thus, we apply the backbone extraction algorithm explained in the method section on all generated graphs. The algorithm removes all noisy edges from the input graph and retains only relevant once. Figure~\ref{fig:coocc} plots the backbone networks of each topic. We use Gephi\footnote{\url{http://gephi.org}, last accessed on Feb, 2016} tool to draw the graphs. Colors in the graphs correspond to the different communities of hashtags identified by Gephi's internal modularity detection algorithms. The first graph (Global) is generated using all the climate change related hashtags. We clearly see the existence of four big communities of hashtags around: \#environment, \#climatechange, \#qatar. Interestingly, we see that the \#climatechange community includes two heavy nodes: \#cop18 and \#cop21. While \#cop18 is heavily connected to the community around \#qatar just as the other hashtag \#doha (the capital), \#cop21 (which happened this year in Paris) does not seem to be connected to Qatar.  As the other COP editions didn't attract media's attentions comparing to the above-mentioned ones, our study dataset reflects the same pattern. 

The backbone network corresponding to general topic is nicely clustered too. In fact, it has a giant component of hashtags around \#climatechange and a small yet accurate one around \#drought containing hashtags such as: California, Haiti, Somelia, rain, and water. All these hashtags are related in a way or another to drought. The absence of rain and water are the main causes of drought, and the mentioned countries and cities are the ones that are struggling with ongoing drought.   
Negotiations\&Summits network is also very informative. The modularity detection algorithm used was able to partition the network into four main communities: \#cop17, \#cop18, \#cop19, and \#cop21, with a high polarization around \#cop18 (Doha) and \#cop21 (Paris) which are both associated with some distinctive hashtags. Interestingly one can easily see that COP17 (Durban, South Africa), COP19 (Warsaw, Poland) and COP20 (Lima, Peru) did not attract as much attention as COP18 (Doha, Qatar) and COP21 (Paris, France) if not at all. While COP18 is heavily covered for obvious reasons, we believe that the Paris' COP21 local attention is mainly due to the fact that Paris was already catching the worldwide attention for the sad Paris attacks event that happened two weeks before the start of COP21 \footnote{November 2015 Paris attacks: \url{https://en.wikipedia.org/wiki/November_2015_Paris_attacks}}, thus benefiting from a large international media coverage that was already in place. It is also interesting to see that both conferences are associated with distinctive sets of hashtags. For instance, COP18 is associated with hashtags such as \#climatelegacy, \#kyoto, \#nojusticenodeal, \#ethiopia, and \#philippines whereas COP21 is associated with hashtags such as \#climatejustice, \#carbon, \# renewables, \#climatefinance, \#energy, \#morocco, and \#china. Interestingly, the hashtag \#Aljazeera which refers to the name of the famous Qatar-based media network appears on the COP21 side. 
Politics\&Opinion backbone network presents an interesting topology with a strong emphasize on \#environment which is linked to three smaller communities: \#qatar, \#climate, and \#waste (represented in blue). 
The remaining graphs are self-explanatory. However, it is worth notice that the relatively small networks of Oceans\&Water and Risk\&Disaster have only one community. Arctic and Air Quality networks are not represented here because the former returned an empty backbone graph whereas the latter had only two nodes, namely: \#airpollution and \#health. 

\subsection{Environment-aware users}
\begin{figure}
\centering
\subfigure[Top 1\% most prolific users.\label{fig:top1_users}]
{\includegraphics[width=.48\linewidth]{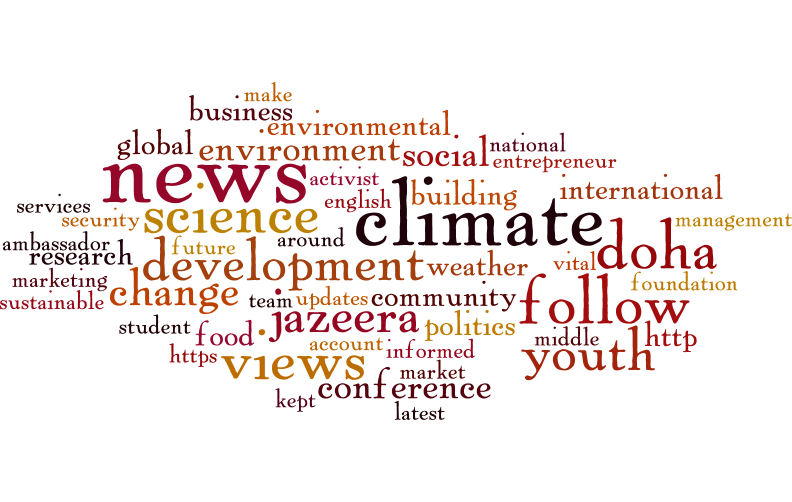}}
\hfill
\subfigure[Top 5\% most prolific users.\label{fig:top5_users}]
{\includegraphics[width=.48\linewidth]{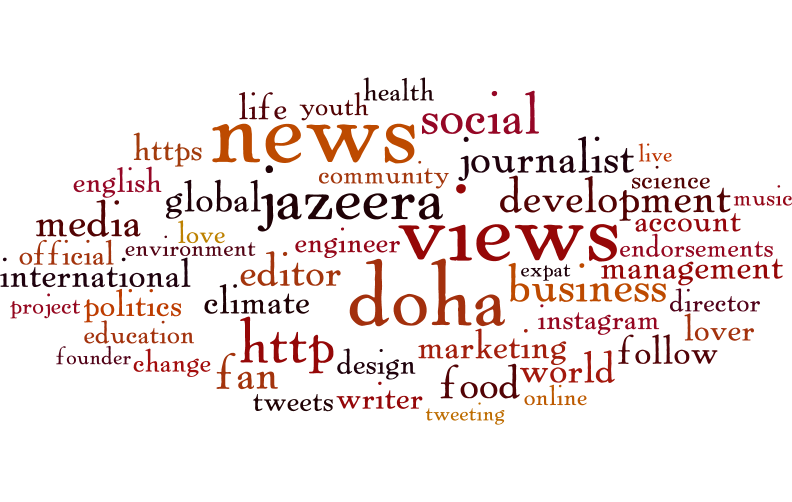}}
\subfigure[All users.\label{fig:all_users}]
{\includegraphics[width=.48\linewidth]{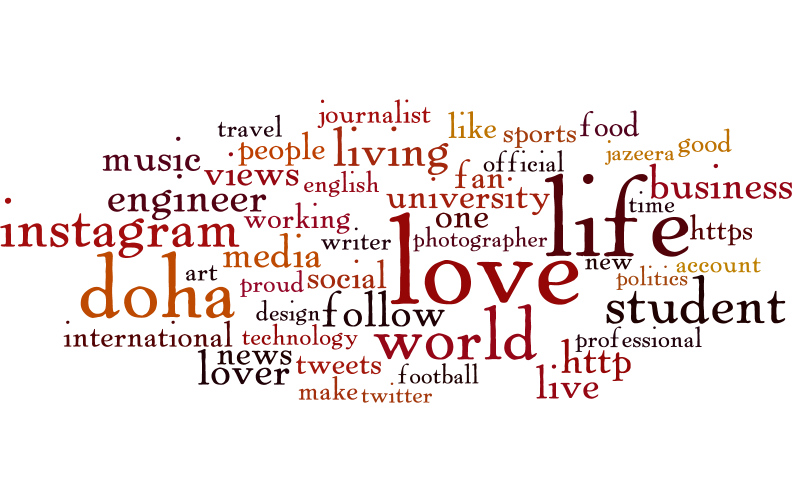}}
\hfill
\subfigure[Complementary cumulative distribution function.\label{fig:ccdf}]
{\includegraphics[width=.48\linewidth]{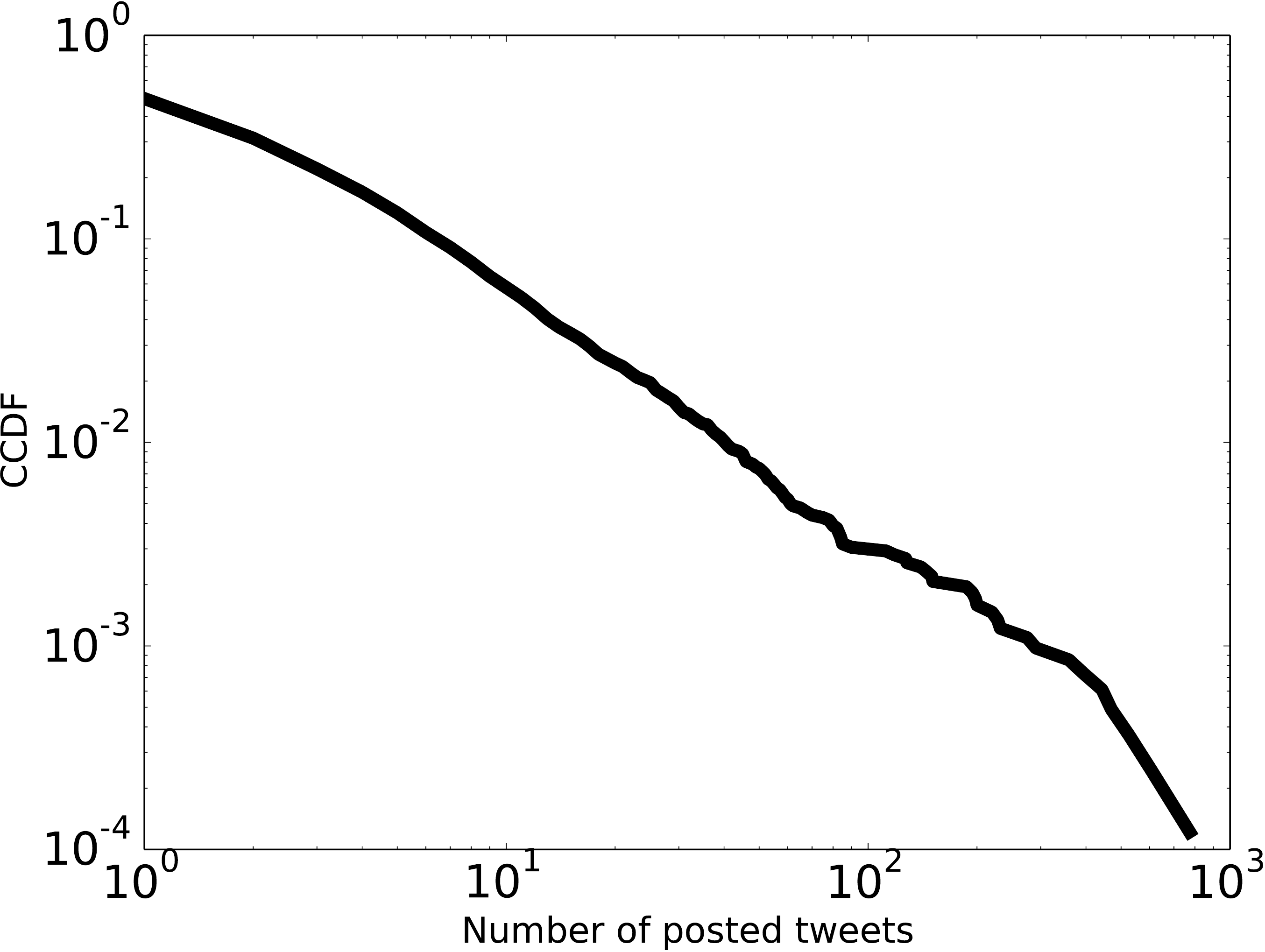}}
\caption{Results for users analysis. (a) through (c) represent word clouds of terms found in user biographies retrieved from Twitter. We see that the more a user tweets about climate change, the more likely she is to have climate change related terms in her profile. (d) shows the skewed distribution of number of climate change related posts per users, with more than half users posting only once.}
\label{fig:users}
\end{figure}

After our content-centered analysis, we work on analyzing users. We use \texttt{Users} endpoint of Twitter API to request the profiles of  8K users who posted at least one tweet about environment. Each user profile comes with basic information such as {\em id, screen name, full name, bio} and some statistics about the number of {\em posts, followers}, and {\em friends}. We analyze the profiles of users, particularly their self-identified biographies. Notice that while not all users provide a biography in their profiles, most of them do. Our objective is to get insights into the interests of users who tweet about environment and climate change. Thus, we merged all users' biographies into one file and perform some cleaning tasks such as the removal of Arabic and English stop words.  
We show in Figure~\ref{fig:users} the most frequent 50 terms found in the biographies of top 1\% (a), top 5\% (b) most tweeting users, and that of all users (c). 
Unsurprisingly, the more users who tweets about climate change, the more likely they mention a climate change related term in their biographies. In other words, this means that prolific users in our dataset (top 1\%) are definitely environmental activists who identify themselves with words such as: climate, change, development, environment and  sustainable. The second tier of users (top 5\%), present a media and news oriented profiles. These users mention in their biographies words such as: news, views, journalist, aljazeera (which is one of the biggest news groups in the world). When all the users are taken together, all signals referring to climate change fade out. In fact, at scale, we found that more than half of user users (51.08\%) have tweeted only once about climate change in a course of 5 years (2011 - 2015). These users tend to mention broader topics in their biographies such as: love, life, world, etc. Notice that the word Qatar has been removed from all three word clouds as it dominated all of them. Figure~\ref{fig:ccdf} plots the complementary cumulative distribution function for the probability (y-axis, log scale) that a random user would post more than a given number of tweets (x-axis). The figure shows a rapidly decreasing trends with most users posting only one tweet, and very few (p=0.005) who posted a hundred tweets or more.

\begin{figure}[th]
  \centering
  \includegraphics[width=0.99\columnwidth]{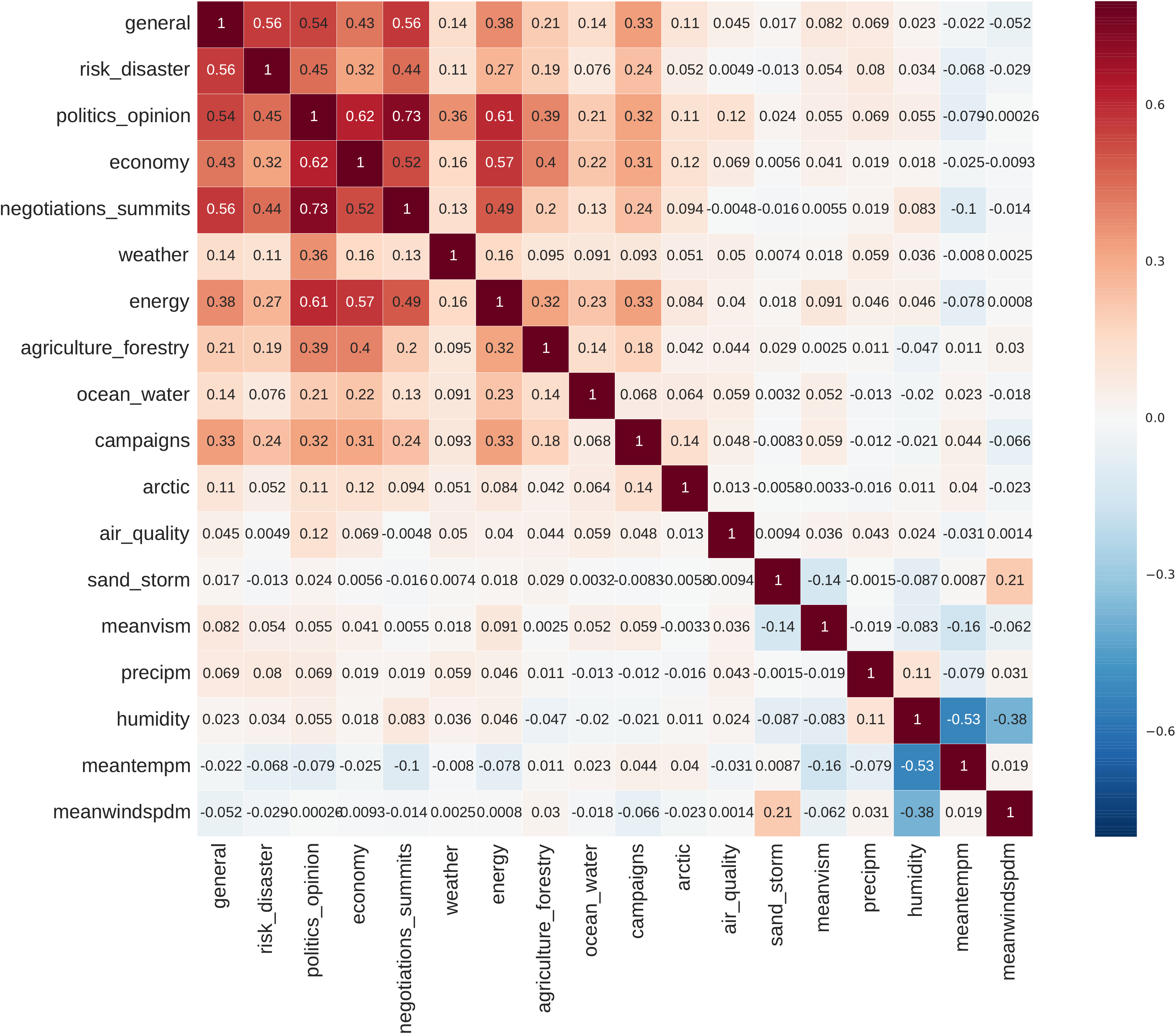}
  \caption{Matrix of pairwise correlations between different climate change and weather variables. The highest {\em p-value} score observed in this matrix is $4.64e-07$ which means that all reported correlations are statistically significant.}
  
  \label{fig:cocc_matrix}
\end{figure}

\subsection{Correlations}
The correlation analysis performed on the 18 variables (11 UN climate change topics, 2 Qatar specific topics, and 5 weather measurements) revealed the existence of a significant dependency between some of them. 
Figure~\ref{fig:cocc_matrix} shows the Pearson correlation score computed over the daily time series of all the pairs.

Looking solely at correlations between climate change topics, we find that Politics\&Opinion and Negotiations\&Summits achieve the highest correlation score of $\rho=0.73$. This definitely makes sense as many political activities and ``pourparlers'' take place during the negotiations. The second and third highest correlation scores are achieved between Politics\&Opinion and Economy ($\rho=0.62$) and Politics\&Opinion and Energy ($\rho=0.61$). This is also expected as politics is tightly related to the two aforementioned topics. Overall, one could easily see that the topics General, Negotiations\&Summits, Politics\&Opinion, Economy, and Energy constitute a well correlated group. Expending our analysis to include the weather variables, we find that Sandstorm topic showed some significant positive dependencies with mean wind speed ($\rho=0.21$) and a negative correlation with visibility ($\rho=-0.14$). Thus, the lower the visibility, the more likely a sandstorm is going on, which means more people are tweeting about it. The other low yet statistically significant correlation is observed between Negotiations\&Summits and mean temperature ($\rho=-0.1$) which is simply explained by the fact that the most important summits captured in our collection (COP18, COP21) took place in winter time. 

\section{Conclusions and Future Work}

In our study, we start by monitoring, collecting, filtering, and analyzing a large dataset of 109 millions tweets posted by 98K distinct users, in a course of 9 years in order to capture the public interest in climate change. We use a taxonomy of climate change to classify a significant amount of relevant tweets into different topics covering a wide range of issues such as politics, economy, and air quality. We perform different types of analysis to understand and reveal the timely distribution of public interest toward different topics.

In our examination of the relevant sub set of tweets, the results revealed that people's interest is mainly driven by widely covered events (e.g., Paris COP21), or local events that have a direct impact on users' daily life (e.g., sandstorms, Doha COP18). The number of users who engage in climate change discourse does not seem to be on a increasing trend. Our analysis revealed that the burst in users' activity observed during COP18 has quickly fade out after the conference, suggesting that organizing big events is not enough to raise any lasting public awareness toward climate change. The backbone graph analysis on the other side, has shown that hashtag co-occurrence is not random and leads in most of the cases to meaningful connected communities (clusters of hashtags).

The findings of this paper did not just highlight interesting answers on public's opinion about climate change in Qatar, but also raise questions for future research to explore. First, we would like to alleviate the limitation of keyword based filtering by designing a better classifier that accounts for semantics of tweets. Second, we aspire to complement this work with a longitudinal analysis of causality between different climate change topics using transfer of entropy technique. Finally, we would like to extend this study to other countries in the region and examine the geographical influences on public's opinion about climate change. 

\balance
\bibliographystyle{aaai}
{\small
\bibliography{biblio}}

\end{document}